\documentstyle[preprint,prl,aps]{revtex}

\begin{document}
\renewcommand{\thefootnote}{\fnsymbol{footnote}}
\begin{flushright}
{\small
SLAC--PUB--7413\\
LBNL--40054\\
February 1997\\}
\end{flushright}

\begin{center}
{\bf\large   
Bremsstrahlung Suppression due to the LPM and Dielectric Effects in
a Variety of Materials\footnote{Work supported by
Department of Energy contract  DE--AC03--76SF00515.}}

\smallskip

P.L.~Anthony,$^1$ 
R.~Becker-Szendy,$^1$
P.~E.~Bosted,$^2$ 
M.~Cavalli-Sforza,$^{3,\#}$ 
L.~P.~Keller,$^1$ \\
L.~A.~Kelley,$^3$ 
S.~R.~Klein,$^{3,4}$ 
G.~Niemi,$^1$ 
M.~L.~Perl,$^1$
L.~S.~Rochester,$^1$ 
J.~L.~White$^{1,2}$
\medskip

{$^{1}$Stanford Linear Accelerator Center,
Stanford, CA 94309} \\
{$^{2}$The American University, Washington, D.C. 20016}  \\
{$^{3}$Santa Cruz Institute for Particle Physics,
University of California, Santa Cruz, CA   95064} \\
{$^{4}$Lawrence  Berkeley Laboratory, Berkeley, CA 94720} \\
\end{center}

\vfill

\begin{center}
{\bf\large   
Abstract }
\end{center}

The cross section for bremsstrahlung from highly relativistic
particles is suppressed due to interference caused by multiple
scattering in dense media, and due to photon interactions with the
electrons in all materials.  We present here a detailed study of
bremsstrahlung production of 200 keV to 500 MeV photons from 8 and
25~GeV electrons traversing a variety of target materials.  For most
targets, we observe the expected suppressions to a good accuracy.  We
observe that finite thickness effects are important for thin targets.

\vfill

\begin{center}
Submitted to {\it Phys. Rev. D}
\end{center}

\centerline{PACS  Numbers: 13.40.-f,12.20.-m,41.90.+e,42.50.Ct}

\vfill \newpage

\narrowtext
\section{Introduction}

When an ultra-relativistic electron emits a low energy photon via
bremsstrahlung, the longitudinal momentum transfer between the
electron and the target nucleus can be very small.  Because of the
uncertainty principle, this means that the momentum transfer must take
place over a long distance, known as the formation length.  One way to
think of this is as the distance required for the electron and photon
to separate enough to be considered separate particles.

If anything happens to the electron or photon while travelling this
distance, the emission can be disrupted.  We have previously presented
letters demonstrating suppression due to multiple scattering\cite{prl}
and dielectric suppression\cite{dprl}.  We present here additional
data further exploring these suppression mechanisms in a variety of
materials.  These data explore bremsstrahlung production of 200 keV to
500 MeV photons from 8 and 25~GeV electrons.  Special attention
will be given to the effects of finite target thickness.

\subsection{LPM Suppression}

LPM suppression is due to multiple scattering, first discussed by
Landau and Pomeranchuk\cite{landau} and slightly later by
Migdal\cite{migdal}.  If an electron multiple scatters while
traversing the formation zone, the bremsstrahlung amplitude from
before and after the scattering can interfere, reducing the amplitude
for bremsstrahlung photon emission.  A similar suppression occurs for
pair production.

The LPM effect is relevant in many areas of physics.  It will cause
the elongation of high energy electromagnetic showers, making them
appear more like hadronic showers.  At the next generation of
colliders, LHC and NLC, this may reduce the electron-pion separation
achievable for a given detector configuration, especially where early
shower development is monitored with a pre-shower detector.

The effects of LPM suppression on cosmic ray air showers have been
discussed by many authors\cite{cosmics}.  In exceedingly high energy
(above $10^{18}$ eV) photon induced air showers, the LPM effect
increases the graininess of the shower, and changes the relationship
between shower density and calculated energy.  LPM suppression can
also affect showers produced by ultra-high energy $\nu_e$ interactions
in water or ice, as might be observed by underwater or under-ice
detectors\cite{DUMAND}.

The electronic LPM effect has analogs in nuclear physics involving
quarks and gluons moving through matter, and calculations have used
LPM-like formalisms to put limits on color dE/dx\cite{Brod1}.
However, the strong-coupling nature of QCD makes comparison with data
less than straightforward. An LPM-type suppression also appears is in
stellar interiors. Because the density is very high, the nucleon
collision rate, $\Gamma_{coll}$, far exceeds the oscillation frequency
of neutrino or axion radiation\cite{Raffelt}, production of these
exotic particles is suppressed.

Several previous experiments have studied the LPM effect, mostly with
cosmic rays.  Most of the cosmic ray experiments date to the
1950's\cite{Fowler}, with a few more recent
results\cite{Strausz}. Most examined the depth of pair conversion of
high energy photons in emulsion.  They qualitatively confirmed the LPM
effect, but with very limited statistics.

A 1975 experiment at Serpukhov measured the photon spectrum from
40~GeV electrons\cite{Serp}. They were troubled by limited statistics
and large systematic errors and backgrounds, but observed a
qualitative agreement with the LPM theory.  Experiment CERN NA-43
measured photon emission from electrons and positrons in a silicon
crystal\cite{bak}.  They observed suppression due to a number of
effects; they attribute part of the total to the LPM effect.
 
\subsection{Dielectric Suppression}

A second suppression mechanism involves the photons.  Produced photons
can interact with the electrons in the medium by Compton scattering.
For forward scattering, this interaction can be coherent, causing a
phase shift in the photon wave function.  If this phase shift, taken
over the formation length, is large enough, then it can cause a loss
of coherence, reducing photon emission.  As the photon energy
approaches zero, this effect completely suppresses bremsstrahlung,
removing the infrared divergence of the original Bethe-Heitler cross
section.  This is the QED analog of color screening in
QCD\cite{screen}. Little previous data exist on this suppression
mechanism\cite{Armenia}.

\section{Theory}

The length scale for suppression is determined by longitudinal momentum
transfer from the nucleus to the electron: 
\def\pe{$p_e$}
\def\pep{$p_e'$}
\def\eep{E$_e'$}
\def\qpar{q$_\parallel$}
\def\thetams{$\theta_{MS}$}
\def\thetab{$\theta_B$}
\def\xo{X$_0$\ }
\begin{equation}
q_\parallel = p_e - p_e' - k = \sqrt{E^2 - m^2} -\sqrt{(E-k)^2 -
m^2} - k,
\end{equation}
where \pe ~and $E$ are the electron momentum
and energy before the interaction, \pep\ is the electron
momentum afterward, $m$ is the electron mass, and $k$ is the
photon energy.  For $E\gg m$ and $k\ll E$, this
simplifies to 
\begin{equation}
q_\parallel \sim {m^2 k \over 2 E (E - k) } \sim
{k \over 2 \gamma^2},
\end{equation}
where $\gamma=E/m$. This momentum can be very small, for example, 0.02
eV/c for a 25~GeV electron emitting a 100 MeV photon. Therefore, the
uncertainty principle requires that the emission take place over a
long distance, called the formation length:  $l_f~=~2\hbar
c\gamma^2/k$.  For 25 and 8~GeV electrons, $l_f (m)~=~864$eV$/k$ and
$l_f (m)~=~88.2$eV$/k$ respectively.  This is the same formation
length that occurs in transition radiation\cite{Jackson}.

\subsection{LPM Suppression}

The LPM effect comes into play when one considers that the electron
must be undisturbed while it traverses the formation length.  One factor that
can disturb the electron, and suppress the bremsstrahlung, is multiple
Coulomb scattering.  If the electron multiple scatters by an angle
\thetams, greater than the typical emission angle of bremsstrahlung
photons, \thetab $\sim$ m/E$=1/\gamma$, then the bremsstrahlung is
suppressed.
 
In the Gaussian approximation, a particle traversing a thickness
$l_f$ of material with radiation length $X_0$ scatters by an
average angle of\cite{Rossi}
\begin{equation}
\overline\theta_{MS}^2 = ( {E_s \over E} )^2 {l_f \over X_0},
\end{equation}
where $E_s=\sqrt{4\pi/\alpha}\cdot m = 21$~MeV and
$\alpha$ the fine structure constant $\sim 1/137$.
\def\elpm{E$_{LPM}$}
The LPM effect becomes important when $\theta_{MS}$ is larger than
$\theta_B$.  This occurs for $ E_s/E \sqrt{l_f/X_0} > m/E$.  For a
given electron energy, suppression becomes significant for photon
energies below a certain value, given by 
\begin{equation}
y = {k \over E} < {E \over E_{LPM}},
\end{equation} 
where \elpm (eV) = $m^4 X_0 / (2\hbar c E_s^2$) = $3.8\times10^{12}
X_0$(cm), about 1.3~TeV in uranium and 2.1~TeV in lead; values for the
targets used in this experiment are given in Table~I.  For a specific
beam energy, 25~GeV, for example, it is possible to define a maximum
photon energy for which the LPM effect is significant, $k_{LPM} =
E^2/E_{LPM}$ For example, $k_{LPM25}$ is 470~MeV for uranium, and 8.5
MeV for carbon; Table~I gives values for our targets for 8 and 25~GeV
beams.

The multiple scattering adds to $q_\parallel$ by changing the
electrons direction, and reducing its momentum.  The formation zone
can be found by replacing $p$ and $p'$ with their forward components
assuming that the multiple scattering is spread throughout the
formation zone.  Then,
\begin{equation}
q_\parallel = ({k\over 2\gamma^2})(1  + {E_s^2 l_f \over 2 E^2 X_0}).
\end{equation}
Since the formation zone length is given by
$l_f=\hbar/q_{\parallel}$, this produces a quadratic equation for
$l_f$, and, hence suppression:
\begin{equation}
S = \sqrt{k E_{LPM} \over E^2}.
\end{equation}
Migdal did a detailed calculation, describing the multiple scattering
angles classically with a Gaussian distribution, and solving the
transport equation to find an ensemble of trajectories\cite{migdal}.
Then, with appropriate weighting, he used these trajectories to
calculate the photon emission probability.  He found 
\begin{equation}
  {d\sigma_{\text{LPM}} \over dk} = {4\alpha r_e^2\xi(s) \over 3k}
\{y^2 G(s) + 2 [1 +(1-y)^2 ] \phi (s) \} Z^2  \ln\bigg( {184 \over Z^{1/3}}
\bigg),
\end{equation}
where
\begin{equation}
s = {1\over 2} \bigg( {y \over 1-y} \bigg)^{1/2} \bigg( {mc \over \hbar}
{mc^2 \over E} {\alpha X_0 \over 8\pi \xi(s)} \bigg)^{1/2}.
\end{equation}
Z is the atomic number, $r_e$ the classical electron radius, and
$\xi(s)$, G(s) and $\phi(s)$ are complex functions with
$1\le\xi(s)\le2$, $0\le G(s) \le 1$ and $0\le\phi(s)\le1$.  When $y\ll
1$, $s\sim\sqrt{(k E_{\text{LPM}}/E^2)}$.
In the absence of suppression
$s\rightarrow\infty$, $G(s)\rightarrow 1$, and $\phi(s) \rightarrow1$;
strong suppression corresponds to $s\rightarrow0$, $G(s)\rightarrow
0$, and $\phi(s) \rightarrow0$. Migdal's calculation gives results 
within about 10\% of Eqn.~6.

Migdal was forced to made a number of simplifying assumptions. First,
he only included elastic scattering from the nuclei themselves.  More
recent calculations have considered both electron-nucleus and
electron-electron interactions, using form
factors\cite{perl}\cite{Tsai}:
\begin{equation}
{d\sigma_{\text{BH}} \over dk} = {4\alpha r_e^2 \over 3k}\bigg[ \{ y^2
    + 2 [1 + (1-y)^2] \} (Z^2 F_{el} + Z F_{inel})+ (1 - y) {( Z^2 + Z
    )\over 3}\bigg].
\end{equation}

Here $F_{el}\approx \ln( 184 / Z^{1/3})$ and $F_{inel}\approx \ln(
1194 / Z^{2/3})$ are the elastic and inelastic atomic form
factors\cite{Tsai}.  In Eqn.~7, $d\sigma_{\text{LPM}} / dk$ includes
the elastic form factor, but not the inelastic form factor or the last
$(1-y)(Z^2+Z)/3$ term. Because the elastic and inelastic form factors
have the same $y$ dependence, it is easy to include the inelastic form
factor by normalizing $d\sigma_{\text{LPM}}/dk$ to the radiation
length as defined by Tsai\cite{Tsai}.  Because of the small momentum
transfer, the recoil of the struck electron can be neglected, and so
electron-electron bremsstrahlung should manifest the same LPM
suppression as nuclear bremsstrahlung.

The $(1-y)(Z^2 + Z)/3$ term is omitted from both our cross sections
and the traditional definition of the radiation length\cite{Tsai};
this is roughly a 2\% correction.

In addition, Migdal was forced to assume that the multiple scattering
angle followed a Gaussian distribution; this is known to underestimate
the number of large angle scatters.  This can affect his results.  For
example, the occasional large angle scatter can lead to some
suppression at photon energies above which Migdal predicted
suppression would disappear.

Blankenbecler and Drell developed a new calculational approach to this
suppression, based on the formalism they developed for beamstrahlung,
treating the multiple scattering quantum mechanically\cite{Drell}.
The results of their calculation cannot be given as a simple equation,
but their results are similar to those of Migdal for thick targets.  

One big advantage of their calculation is that it implicitly handles
targets of finite thickness, dividing the electron path into 3
sections: before the target, inside the target, and after the target,
with interference between the different regions (including before
and after). Because of this treatment, they calculate the total emission
over the slab, and do not localize the point of photon emission.

More recently, Zakharov has presented a
calculation\cite{Zakharov}.  Although it has a different basis from
Blankenbecler and Drell, it appears to give similar results.
Unfortunately, it also suffers from the same limitations regarding
multiple emission and dielectric suppression.

\subsection{Dielectric Suppression}

The magnitude of dielectric suppression, due to the photon-electron
gas interactions, can be calculated by finding the photon phase shift
due to the dielectric constant of the medium, using classical
electromagnetic theory\cite{mikaelian}.  The phase shift is
$(1-\sqrt{\epsilon})kcl_f$ where $\epsilon$ is the dielectric constant
of the medium, given by
\begin{equation}
\epsilon(k)=1-(\hbar\omega_p)^2/k^2,
\end{equation}
where $\omega_p=\sqrt{4\pi N Z e^2/m}$; N is the number of atoms per
unit volume, Z the atomic charge, and $e$ the electric charge.  If the
phase shift gets large, coherence is lost.  This limits the
effective formation length to the distance which has a phase shift of
1:
\begin{equation}
l_f={2\hbar c k \gamma^2 \over k^2+ k_p^2},
\end{equation}
where $k_p=\gamma\hbar\omega_p$ is the maximum photon energy for which
dielectric suppression is large.  It is also the maximum energy
at which transition radiation is large.
The suppression is simply given by the ratio of in-material to
vacuum formation lengths:
\begin{equation}
S = { k^2 \over k^2 + k_p^2}.
\end{equation}
The suppression becomes large for $k < k_p$; below this energy, the
photon spectrum changes from $1/k$ to $k$.  Numerically, the plasma
frequencies for most solids are in the 20--60 eV range, so the
suppression becomes important for $k < r E$ where $r=\hbar\omega_p/m=
\hbar\sqrt{4\pi Z e^2/m^3}$, about $5.5\times10^{-5}$ in carbon or
$1.4\times10^{-4}$ in tungsten; values for other targets are given in
Table~I.  For small $k$, dielectric suppression is much more important
than LPM suppression.

\subsection{Total Suppression}

Because LPM and dielectric suppression both reduce the effective
formation length, the suppressions do not simply multiply.  Where both
mechanisms appear, the total suppression can be found by summing the
contributions to $q_\parallel$ and hence $l_f=\hbar/q_\parallel$; the
suppression is simply the ratio of $l_f$ to its vacuum
value\cite{galitsky}. Migdal included dielectric suppression in his
formalism by scaling $\phi$ appropriately\cite{migdal}.
Unfortunately, the Blankenbecler and Drell approach is not easily
amenable to inclusion of dielectric suppression\cite{dick}.

For 25~GeV beams hitting the targets used here, the LPM effect is more
important for photon energies above 5 MeV; at significantly lower
energies, dielectric suppression dominates.  With 8~GeV beams, LPM
suppression is reduced by a factor of $(8/25)^2$, so dielectric
suppression is usually the dominant effect. These spectral shape for
the different photon energies (and hence mechanisms) are schematically
summarized in Fig.~1.

\subsection{Thin Targets and Surface Effects}

When an electron interacts near the surface of a target, part of the
formation zone may extend outside of the target. Then, there will be less
multiple scattering or Compton scattering, so the suppression should
be reduced.  There is also a transition as the electromagnetic fields
of the electron readjust themselves to allow for the electron multiple
scattering and effect of the medium.

A very  simplistic approximation for the surface effects would be
to allow for a single formation length of un-suppressed Bethe-Heitler
radiation near the target surfaces, with the rest of the radiation
from the interior fully suppressed.  This implies that the surface
effects are important where LPM suppression is large, at small $k$,
since $l_f$ scales as $1/\sqrt{k}$. However, where dielectric
suppression dominates, $l_f$ scales as $k$, giving short formation
zones and little surface effects.

Unfortunately, this model is conceptually inadequate because, in
addition to the reduced suppression, there can also be edge radiation.
For dielectric suppression, this is just conventional transition
radiation\cite{Jackson}, given by
\begin{equation}
{dN \over dk }= {\alpha \over \pi k} \bigg[(1+{2k^2 \over k_p^2})
\ln{(1+{k_p^2 \over k^2})} -2 \bigg].
\end{equation}
Where LPM suppression is large, Gol'dman\cite{goldman} has pointed out
that there is an additional transition radiation caused by the
multiple scattering.

When the target is thinner than the formation zone, the problem
simplifies.  For extremely thin targets, where the target thickness
$t< X_0 (m/E_s)^2$, there isn't enough multiple scattering to 
cause suppression, and the Bethe-Heitler spectrum is retained.

For slightly thinner targets, but where $t< l_f$, Shul'ga and Fomin
showed\cite{Shulga} that the entire target can be treated as a single
radiator, and the Bethe- Heitler spectrum is
recovered\cite{ternovskii}, albeit at a reduced intensity.  The
radiation spectrum is given by
\begin{equation}
{dN_{SF}\over dk} = {2\alpha\over\pi} \int_0^\infty
d^2\theta f(\theta) \bigg( {2\zeta^2+1 \over \zeta\sqrt{\zeta^2+1}}
\ln{(\zeta+\sqrt{\zeta^2+1})}-1\bigg),
\end{equation}

where $\zeta=\gamma\theta/2$, $\theta$ being the scattering angle.
The
integrals are taken over the two independent scattering planes, and

\begin{equation}
f(\theta)={1\over \pi\theta_0^2} \exp(-\theta^2/\theta_0^2).
\end{equation}
Because the targets are very thin\cite{PDG},
\begin{equation}
\theta_0= {E_s \over E} \sqrt{t\over X_0}\  [1+0.038 \ln{t\over X_0}].
\end{equation}
These formulae are numerically evaluated.  It is worth pointing out
that, in the limiting case, the radiation becomes proportional to
ln($t$)!  Then, the radiation depends only on $t/X_0$, and is
independent of $E$.  This spectrum applies for photon energies $k$
where the reduced formation length (taking into account the reduction
due to the LPM effect) is larger than the target thickness.  This
occurs when
\begin{equation}
l_f = S*l_{f0} =  {dN_{SF}/dk \over dN_{BH}/dk } 
{2\hbar c\gamma^2\over k} > t,
\end{equation}
where $dN_{BH}/dk$ is the Bethe-Heitler predicted radiation from the
entire sample.  This equation is valid as long as dielectric
suppression and transition radiation are not large.

For thicker targets, Ternovskii\cite{ternovskii} calculated the
spectrum of this radiation at an interface.  Like Blankenbecler and
Drell, Ternovskii divided the electron path into 3 regions, and allowed
for interference between the regions.  For sufficiently thick targets,
he parameterized his results into a bulk emission, matching Migdal,
plus two edge terms. For $k\ll E$ and $s\gg 1$, the edge term is
conventional transition radiation.  For $s<1$ and $sk_p^2/k^2\ll 1$ LPM
suppression dominates and Ternovskii finds for $k\ll E$,
\begin{equation}
{dN \over dk}= {2 \alpha \over \pi k} \ln{{\chi\over \sqrt{s}}},
\end{equation}
where $\chi\sim 1$, similar to the logarithmic uncertainty found by
Migdal. For $s>1$, the region of no LPM suppression, this equation is
negative; common sense seems to indicate that the function should be
cut off. For comparison with data, a more serious problem is that
Eqns.~13 and 18 do not match up in the region $sk_p^2/k^2\sim 1$.

Garibyan\cite{garibyan} also calculated the transition radiation
spectrum, also using Gol'dman as a base,  but for a single edge.  His
results were similar, but not identical to Ternovskii, with the same
negative region.

In 1965, Pafomov \cite{pafomov} stated that the formulations of
Gol'dman, Ternovskii and Garibyan were flawed because they improperly
separated the total radiation into bremsstrahlung and transition
radiation, causing the negative regions.  In his calculations, Pafomov
found that there is transition radiation even for $s>1$, with a
$1/k^2$ spectrum.  Pafomov predicted that, for $k_{LPM} > k_p$,
the transition radiation term is, per edge:

\begin{eqnarray}
{dN \over dk} = {\alpha\over \pi k } \log{(k_p/k)^2}, &\hskip .4 in
k < k_p^{4/3}/k_{LPM}^{1/3} \\
{dN \over dk} = {\alpha\over \pi k} \log{{2 \over 3} \sqrt{{k_{LPM}\over k}}}, 
& \hskip .4 in k_p^{4/3}/k_{LPM}^{1/3}< k \ll k_{LPM} \\
{dN \over dk} = {8\alpha\over 21\pi k} ({k_{LPM} \over k})^2.  &\hskip .4 in
k \gg  k_{LPM}
\end{eqnarray}

The first equation is similar to, but larger than conventional
transition radiation, with the difference probably due to
calculational technique.  Unfortunately, Eqns.~19 and 20 do not quite
match when $k= k_p^{4/3}/k_{LPM}^{1/3}$ causing a noticeable step in
our simulations.  There is also a discontinuity between Eqns.~20 and
21 at $k\sim k_{LPM}$.  Pafomov gives a numerical approximation
that covers the entire region $k>k_{LPM}^{4/3}/k_p^{1/3}$ and avoids
the discontinuity; we use it in our calculations.
For bulk emission, Pafomov accepted Migdal's results.

Because of the logarithmic uncertainties, transition regions, and
discontinuities, it is difficult to confidently apply any of these
edge effect formulae; we will show a few selected comparisons with our
data.  Even in the absence of a acceptable theory, it is possible to
remove the edge effects by comparing data from targets of similar
composition but different thickness.  By subtracting the two spectra,
it is possible to find an `internal' spectrum and a `surface effect'
spectrum, accurate as long as there is no interference between the two
edge regions.

For thin targets, dielectric suppression should be reduced, at least
in classical calculations.  When the photon wave phase shift, integrated
over the target thickness, is small, then the suppression should
disappear.

\section{Experiment}

This experiment\cite{prl} \cite{dprl} \cite{klein}\cite{becker} was
conducted in End Station A at the Stanford Linear Accelerator
Center. As Fig.~2 shows, electrons entered End Station A and passed
through targets mounted in a seven position target holder.  During
data taking, we rotated through the targets, taking $\sim$2 hours of
data on each target.  We took a total of 8 hours of data on most
target/beam energy/calorimeter gain setting combinations.  The targets
materials and thicknesses are given in Table~II; a selection of high
and low Z targets were used, usually with two target thicknesses per
material. Rotations included one position on the target ladder which
was left empty for no-target running to monitor beam related
background.  A 1 cm square silicon photodiode was mounted in another
position.  By measuring the rates of lead glass hits to Si photodiode
hits, we could check for changes in the beam size; the beam position
and shape proved stable with time.

After passing through the targets, the electrons entered an 18D72
dipole magnet, which was run at 3.25 (1.04) T-m of bending for 25
(8)~GeV electrons.  This field bent full-energy electrons downward by
39 mrad; lower energy electrons were bent more.  One especially useful
feature of the magnet was its large fringe field.  Because of this
fringe field, the electron bending started slowly, so synchrotron
photons produced during the initial bending had low momenta; this
reduced the synchrotron radiation background observed in the
calorimeter significantly.  Synchrotron radiation emitted by an
electron pointing at the bottom edge of the calorimeter had a 280 keV
(9 keV) critical energy at 25 (8) GeV.  The average energy deposition
in the calorimeter was 40 keV and 400 eV respectively.

After bending, the electrons exited the vacuum chamber, travelled 15
meters through a helium bag, into 6 planes of proportional wire
chambers\cite{wires}, with a 20 cm separation, arranged Y U Y V Y U
where Y plane wires were horizontal, and U and V planes were were at a
30 degree angle from horizontal, to provide left-right information.
The Y (U/V) planes had a 2 (4) mm wire pitch.  Due to an unfortuitous
choice of angle, the wire chambers had a momentum resolution only
slightly better than a single plane, giving resolution of roughly 90
MeV at 25~GeV.

The electrons were absorbed in a stack of three 10 cm by 10 cm lead
glass blocks, arranged so that full energy electrons hit the middle of
the top block.  This enabled us to accurately count electrons
calorimetrically.  Electrons with energies below 17.4 (5.8)~GeV for 25
(8)~GeV beams missed the blocks and were not counted.  The fraction
not counted was estimated with the Monte Carlo, and was typically
about 1\% per 1\% of $X_0$ target thickness.

Photons produced in the target travelled 50 meters downstream through
vacuum into a BGO calorimeter.  The calorimeter consisted of 45 (a 7
by 7 array with the corners missing) BGO crystals, each measuring 2 cm
square by 20 cm (18 $X_0$) deep\cite{calorimeter}. Each crystal was
read out by a Hamamatsu R1213 1/2'' photomultiplier tube (PMT) with a
linear base. The PMTs detected about 1 photoelectron per 30 keV of
energy deposition in the BGO.  During much of the running, one crystal
in the outermost row was not functional.  The calorimeter was built
and extensively characterized in 1984 as a prototype, and was
reconditioned for this experiment.  In 1984, the nonlinearity in the
100 MeV range was estimated at 2\%; Monte Carlo simulations of leakage
indicate that this does not change significantly at 500 MeV.

The calorimeter was read out by a LeCroy 2282 12 bit ADC.  The ADC
gate was set to 900 nsec, several times the BGO light decay time of
300 nsec.  One advantage of this gate width was that sensitivity
variations due to the 50 nsec time structure of the electron beam were
negligible.  Because the ADC pedestals were known to drift slowly, frequent
pedestal runs were performed.

Calorimeter ADC overflows were detected by histogramming the ADC
output on a channel by channel, run by run basis; the maximum ADC
count was typically ~3950 counts and was easily determined by
inspection.  Events with an ADC overflow were flagged.  

The experiment studied a very wide range of photon energies, from 200
keV to 500 MeV.  This is a considerably wider range than can be
handled by a single PMT gain and ADC, so data were taken at two
different calorimeter gain settings, with the gain adjusted by varying
the PMT high voltage.  The first data set corresponded to 100 keV per
ADC count, and the second to 13 keV per ADC count.  These will
be referred to as `low gain' and `high gain' running respectively.

Initially, a 1/2'' thick scintillator slab was placed in front of the
calorimeter, as a charged particle veto.  When the charged particle
background was found to be small, it was removed.  The only other
material between the target and the calorimeter was a 0.64 mm (0.7\%
$X_0$) aluminum window immediately in front of the calorimeter. This
minimized the number of produced photons that were lost before hitting
the calorimeter.

Scintillator paddles were located above and below the calorimeter.
Their logical AND provided a cosmic ray muon trigger, used to
calibrate the calorimeter.  The paddles could initiate a trigger in
the interval between beam pulses.

Most of the electronics were housed in a single CAMAC crate. Besides
the calorimeter ADC, lead glass block ADCs and wire chamber hit
patterns, we read out a number of additional scintillator paddles on
each beam pulse, irrespective of what happened on that pulse.
Monitoring data, such as the BGO temperature and spectrometer magnet
settings were read out periodically.  We used the
acquisition framework developed by SLAC--E--142/3.

The beams for this experiment were produced parasitically during
Stanford Linear Collider (SLC) operations.  Off axis electrons and
positrons in the SLAC linac struck collimators near the end of the
accelerator\cite{cavalli}.  A useful flux of high energy
bremsstrahlung photons emerged from the edges of these collimators and
travelled down the beampipe, past the bending magnets, and into a
target in the beam switchyard.  This target converted the photons into
$e^+e^-$ pairs, and those electrons within the A-line acceptance angle
were transported to End Station A.

For most of the running, we ran at an average intensity of one
electron per pulse, with the short term averages between 0.8 and 1.5
electrons per pulse as SLC conditions varied.  The average intensity
was changed by adjusting the momentum defining collimators; typical
momentum acceptance was $\Delta P/P\sim 0.2\%$.  The beam optics were
set up so that there was a virtual focus at the calorimeter.  The
typical beam spot vertical and horizontal half widths were 2.5 mm at
25~GeV and somewhat larger at 8~GeV.

\section{calibration}

Since the calorimeter calibration is crucial to experimental accuracy,
several methods were used to calibrate the calorimeter: 400 and
500 MeV electron beams, bremsstrahlung events, and cosmic ray muons.
The calibrations were divided into two classes: relative calibrations,
which were used to measure the relative gain between BGO crystals, and
absolute, which set the overall energy scale.  The most careful 
calibration was done with the `low gain' calorimeter PMT HV setting;
the `high gain' data were calibrated by comparison with the `low gain'
running.

This analysis used the `low gain' data over the range of 5 to 500 MeV.
The `high gain' data are used from 200 keV to 40 MeV.  Between 5 and 40
MeV, the data are combined using a weighted mean.  In this region, the
data agree well; this gives us confidence in our relative
calibrations.

One key factor in the calibration was the BGO temperature, which is
known to affect both the light output and decay time.  We therefore
measured the way that changing temperatures affected the BGO response
to cosmic ray muons, and corrected the data.  The BGO temperature was
monitored by a thermistor throughout the experiment.  The BGO light
output decreased by 2\%/$^0$C, a bit more than other
measurements\cite{Zucc}.  This correction factor was applied to our
data.

The BGO channel gains were controlled by adjusting the PMT high
voltage. Relative high voltages were set with potentiometric dividers,
and the absolute scale was set by two supplies in our counting house.
The relative gains were roughly equalized before the experiment by
normalizing the calorimeter crystal response to 662 keV gamma rays
from a $^{137}$Cs source.  The change from `high gain' to `low gain'
was done by adjusting the voltage on the two supplies.  Since not
every phototube had identical gain vs. voltage characteristics, this
changed the relative gains somewhat.  Because of this, the relative
channel to channel calibrations were done separately for high and low
gain running.

Better measurements of the relative gain came from the cosmic ray data
gathered throughout the run.  The cosmic ray trigger consisted of a
coincidence between the two scintillator paddles bracketing the
calorimeter.  They were placed so that triggers occurred for muons
traversing the center of the BGO.  

The calorimeter absolute energy scale was largely determined with 400
and 500 MeV electron beams.  The electrons were produced
parasitically, as during normal E-146 running.  Because of the low
energy, special precautions were required.  All of the beam line
magnets were degaussed, and the usual power supplies were temporarily
replaced with lower current supplies that could regulate reliably at
the required power levels.  The magnetic fields were monitored with a
flip coil in a magnet that was subjected to identical treatment to the
beam line magnets.  The estimated error on the overall energy scale
calibration is 5\%.

Since the low energy beam had a relatively wide angular distribution, these 
data also provided a check on the crystal to crystal intercalibration.
By examining histograms of reconstructed energy vs. the location where
the electron hit the calorimeter, we estimate that the crystal to
crystal calibration varied by less than 2\%.  Since most of the
bremsstrahlung photons hit the central crystal, this had a negligible
effect on our overall resolution.

For each event, the electron momentum, measured in the wire
chambers, and the photon energy should sum to the beam energy.  Since
the wire chamber energy resolution is determined by geometry, it can
provide an additional check on the calorimeter calibration.
Unfortunately, because of the steeply falling photon spectrum and the
quantization introduced by the wire spacing, this analysis is quite
tricky. However, it analysis confirmed that the calorimeter energy
calibration is good to within 10\%.

The `high gain' data were calibrated by comparison with the low gain
data, mostly using the cosmic rays.  This calibration is accurate to
about 10\%.  

It is worth noting that the calorimeter behavior is significantly
different for the high and low gain data.  At higher energies, the
impinging photons create electromagnetic showers, while at lower
energies, most photons interact via single or multiple Compton
scattering.  Besides the loss in resolution due to the photoelectron
statistics, it is necessary to account for resolution deterioration
because photons can be Compton scattered out the front face of the
calorimeter; the probability of this increases at low energies.  Also,
because of the possibility of a photon Compton scattering twice, in
two widely separated crystals in the calorimeter, the photon cluster
finder loses efficiency; these problems are accounted for in our
systematic errors, which are larger for small photon energies.

\section{Data Analysis}

Because bremsstrahlung is the dominant cross section, event selection
is simple.  Events containing a single electron in the lead glass were
selected.  The calorimeter ADC counts were converted to energy. For
`low gain' running, the total energy observed in the calorimeter was
used directly.  For `high gain' running clustering was required to
remove spurious pedestal fluctuations: we stared with the highest
energy crystal in the event, and added in the energies of all
neighboring crystals that were above the ADC pedestal. 

Because the angular acceptance of the central crystal, 0.2 mrad, was
larger than the typical bremsstrahlung angle, $1/\gamma\sim0.02$\
mrad, even after allowing for the beam divergence, the majority of the
bremsstrahlung photon flux hit the center of the calorimeter, so we
did not have to correct for calorimeter leakage on an event by event
basis. 

Events with a calorimeter energy between 200 keV and 500 MeV were
histogrammed by photon energy, with the bins having a logarithmic
width. The photon intensity, $(1/X_0) (dN/d(log k))=(1/kX_0) (dN/dk)$
is plotted vs. $k$, with $k$ on a logarithmic scale, necessary to
cover the 3 1/2 decades of energy range.  The $y$ axis is chosen so
that the classical Bethe-Heitler $1/k$ spectrum will appear as a flat
line. There are 25 bins per decade of photon energy, giving each bin a
width $\Delta k/k\sim 0.09$,

Although the Bethe Heitler cross section is flat for a logarithmic
energy binning, the corresponding data would not be flat because of
multiphoton pileup.  This is because a single electron traversing
the target may interact twice, emitting two photons.  Because the
photon energies add, this depletes the low energy end of the measured
spectrum and tilts the spectrum.

The logarithmic energy scale and the mismatch between ADC counts and
histogram bin boundaries can create a problem for low $k$.  The uneven
mapping can create a dithering in the histograms, with different
numbers of ADC counts contributing to adjacent bins, creating an
up-down-up pattern, as can be see in Fig.~2 of Ref.~2. To avoid this,
the data below 500 keV were smoothed with a 3 point average with
weights 0.25 : 0.5 : 0.25.  Above 500 keV, the weights of the 2 side
points were reduced logarithmically with the energy, reaching zero at
5 MeV.  

We have previously shown that both LPM and dielectric suppression are
necessary to explain the data; this paper presents a more detailed
examination of the data for a variety of targets. In most cases, only
a single, combined LPM plus dielectric suppression curve is shown.

To produce histograms covering almost 3 1/2 decades of photon energy,
it was necessary to combine data from the high and low gain running.
Above 5~MeV, high gain data were used, while below 40~MeV low gain data
were used.  Between 5 and 40~MeV, weighted averages of both data sets
were used.  Because the agreement between the two data sets was
considerably better than the estimated systematic errors, the actual
combination technique was unimportant.  One run of 0.7\% $X_0$ Au
8~GeV high gain data was removed from the analysis because it was
significantly above both the other high gain data and also the low
gain data.  And, as discussed below, the 0.1\% $X_0$ gold data were
not always consistent.  In all other cases, the data from individual
runs were consistent.

\subsection{Monte Carlo}

A computer code using Monte Carlo integration techniques based on a
set of look-up tables was written to make predictions for the photon
intensity spectra.  This technique was necessary in order to combine
the effects of multiple photon emission from one electron with
predictions for LPM and dielectric suppression and transition
radiation.  Tables of photon production cross sections are generated,
starting with 10 keV photons, with each step in photon energy
increasing exponentially in multiples of 1.02. The Migdal cross
sections are generated using the simplified calculational methods
developed by Stanev and collaborators\cite{stanev}.  Their
parameterizations agree well with Migdal's calculations, without
dielectric suppression. Our calculations include an additional term
for the longitudinal density effect, in the manner prescribed by
Migdal.

A separate table is generated for transition radiation.  This table is
normally filled with conventional transition radiation (Eqn.~13);
the Gol'dman or Pafomov combined formula can also be used.  The
photons from the entry radiation can, of course, interact in the
target. For ease of extrapolation, these tables are then converted to
integral and total cross sections.

The Monte Carlo then begins generating events.  Each electron enters
the target, and entry radiation may be generated. The electron is tracked
through the target in small steps.  The step size is limited so that
the probability of emission at each step is less than 1\%; at most one
photon can be produced per step.  If the electron radiates, the photon
energy is chosen using the integral cross section table.  The photon
energy is subtracted from the electron energy, and the tracking
continues, until it leaves the target, producing another opportunity
for transition radiation.  The possibility of produced photons
interacting in the target by pair production or Compton scattering is
included using another look up table\cite{Overbo}; any photon that
interacted is considered lost.

When one electron emits multiple photons, the photon energies were
summed before histogramming.  The photon energies are then smeared to
match the measured calorimeter resolution.  

In the Monte Carlo curves, at $1.1 < k/k_{LPM} < 1.3$, the LPM curve
rises slightly above the Bethe-Heitler curve.  This rise comes from
Migdal's original equations, because the product $\xi(s)\phi(s)$ can
rise slightly above 1.

The Blankenbecler and Drell theory, as described in Section II.A., does
not allow for the possibility of multiple interactions, and, without
the photon emission point, it isn't easy to include their calculations
in a Monte Carlo, and, consequently, allow for experimental effects,
such as photon absorption in the target.

Because of these problems, in particular the multiple interaction
possibility, we have not implemented their cross sections in our Monte
Carlo.  Instead, we will directly compare their cross sections with
our data, but only for the thinnest targets, where multiple photon emission
is small, and at energies above those where dielectric suppression
occurs.  

\subsection{Backgrounds}
 
Because the calorimeter subtended such a small solid angle,
backgrounds due to photonuclear interactions were small -- only
photons produced with very small $p_\perp$ would hit the calorimeter.

As previously mentioned, the maximum critical energy for synchrotron
radiation from the spectrometer magnet incident on any part of the
calorimeter was 280 keV (40 keV) at 25 (8)~GeV; for synchrotron
radiation hitting the central crystal, the critical energies were much
lower.  Because the synchrotron radiation was painted in a band
downward from the central crystal, it was easy to identify in the
calorimeter. 

For the 25~GeV `high gain' data, synchrotron radiation could be a
significant background.  For the data, backgrounds were reduced with
the cut diagrammed in Fig.~3.  Photon clusters in the lower 25\% of
the calorimeter, below the diagonal lines, were removed.  Photons
reconstructed exactly on the border were kept, but with an appropriate
weighting, 50\% if they were on the border lines, and 75\% at the
center of the center crystal. The data were adjusted upward to
compensate for this 25\% loss of signal.  Because of uncertainty in
the source of after-cut backgrounds, no further corrections are
applied.

The backgrounds were measured with periodic no-target runs.  The
no-target data, both with and without the synchrotron radiation cut
are shown in Fig.~4 for both 8 and 25~GeV running.  Note that this
figure ais normalized as photons per 1000 electrons, whereas
Figs.~5--13 are normalized as photons per electron per radiation
length of the target.  The backgrounds in Fig.~4 can be scaled to the
data in Figs.~5--13 by dividing by 1000 the electron scale factor in
Fig.~4, and again by the radiation length in percent.  At 25~GeV, the
majority of background is synchrotron radiation, which is largely
removed by the cut. At 8~GeV, the cut has little effect; acceptance
corrections occasionally make the post-cut background larger than the
pre-cut.

Except for the region where synchrotron radiation was expected,
backgrounds were always small.  After the cut, backgrounds at 25~GeV
were less than one 200 keV-- 500 MeV photon per 1000 electrons. At 8~GeV,
the background was about a factor of 3 lower, with or without the cut.

\subsection{Discussion of Data}

Figures~5--13 present our data for a variety of target materials,
arranged in order of increasing suppression. For each material, there
is one figure, with four or six panels, showing two target thicknesses
in 8 and 25~GeV beams, plus edge-effect subtracted data (discussed in
Section VI). The 25~GeV `high gain' data have had the synchrotron
radiation removal cut applied.  For lead, there is only one target
thickness.  Occasionally, there are data at only one energy for a
target.  The high gain and low gain calorimeter data have been combined
as previously described; where there are no high or low gain data, the
histogram is cut off at the appropriate energy.

For each target, we compare the data with different Monte Carlo
curves.  Our standard curve, shown by a solid line in all the plots,
is a Monte Carlo including LPM and dielectric suppression, with
conventional transition radiation.  For the thinner targets, we make
comparisons with a number of transition radiation theories.  For these
plots, the Monte Carlo curves have been normalized to match the data,
as discussed in Section VIII.

Figure~5 shows data from the carbon targets. In addition to the
standard Monte Carlo (solid line), LPM suppression only (dotted line)
and a Bethe-Heitler curve (dashed line) are shown for comparison. To
give an idea of the effect of transition radiation, we also show in
Fig.~5a a Bethe-Heitler only curve and in Fig.~5d the suppression
curve, both with no transition radiation, as dot-dashed lines.  The
upturn below about 500 keV for the 25~GeV electron Monte Carlos is
transition radiation (Eqn.~12).  The additional upturn in
the data are consistent with the remaining background. The combined
Monte Carlo does the best job of representing the data.  At 8~GeV, the
suppression is dominated by dielectric suppression; at 25~GeV, the two
effects have a similar magnitude.  At 25~GeV, the suppression appears
to turn on at higher energies and more gradually than predicted by the
Monte Carlo.

Figure~6 shows data from the aluminum targets, with the same three
Monte Carlo curves as in Fig.~5.  The data are slightly below
the Monte Carlo over most of the plot.  Here, the upturn below 500 keV
in the 25~GeV data are consistent with transition radiation plus
remnant synchrotron radiation.  Since the $Z$ of aluminum is twice that
of carbon, the LPM effect is much larger. Because the densities are
similar, dielectric suppression is very similar.  As with carbon, the
LPM effect appears to turn on slightly more gradually than the Monte
Carlo predicts.

Figure~7 shows data from the iron targets, compared with just the
standard Monte Carlo.  The data and Monte Carlo are close, but the
data may have a longer, but more gradual slope than the Monte Carlo
predicts.  Data from the 2\% \xo lead target are shown in Fig.~8, again
with the standard Monte Carlos.

Figure~9 shows data from the tungsten targets.  The fit is quite good
at 8~GeV. At 25~GeV, for $k< 10$ MeV, the data for the 2\% $X_0$
target are above the Monte Carlo.  At 7 MeV, the target thickness is
comparable to the unsuppressed formation length.  Eqn.~17.  shows that
the suppressed length becomes comparable to $t$ below 3.0 MeV.  Below
1.7~MeV, dielectric suppression reduces $l_f$ below $t$.  Between
1.7~MeV and 3.0 MeV, the target should interact as a single radiator;
the straight line on the figure is from Eqn.~14; the height is
significantly above the data.

Figure~10 shows data from the 3\% \xo and 5\% \xo uranium targets.  In
both cases, the 25~GeV data rise above the Monte Carlo at low $k$.
The prediction of Eqn.~14 is shown by the straight line in the 25~GeV
3\% \xo data.  For the 5\% \xo target and the 8~GeV 3\%\xo data, $t >
l_f$ everywhere, so it is appropriate to treat the edge effects in
terms of independent transition radiation.  The transition radiation
predicted by Ternovskii (dotted line) and Pafomov (dashed line) are
shown on these plots, on top of the LPM + dielectric suppression base.

The Ternovskii curve has a jump around 500 keV in the 25~GeV data.
This corresponds to $sk_p^2/k^2=1$, below which transition radiation
from Eqn.~13 applies; the corresponding $k$ is below 200 keV for 8~GeV
electrons.  Below this energy, Ternovskii matches conventional
transition radiation.  Above this energy, Ternovskii predicts a rather
large transition radiation, which does not match the data.  The match
could be improved by adjusting $\chi$.  However, a rather large
adjustment would be required.

Pafomov's predictions jumps at about 800 keV (400 keV), corresponding
to $k=k_p^{4/3}k_{LPM}^{1/3}$.  Below this, his predictions are
considerably above both conventional transition radiation and the
data.  Above the break, the shape looks reasonable, but the amplitude
appears to be a factor of 2 to 3 too big.

Figure~11 shows data from the 6\% \xo and 0.7\% \xo gold targets.  For
the 0.7\% \xo target, the excess flat region extends from about 1 MeV
up to 30 MeV. The downturn for the 0.7\% $X_0$ data above
$k=100$ MeV is due to the natural decrease of the Bethe-Heitler spectrum.

Because the 0.7\% \xo target is thin enough that multi-photon emission
is small, we can compare it directly with predictions that are not
amenable to Monte Carlo simulation.  We do this in Fig.~12, which
shows an enlarged view of the data in Fig.~11.  Here, the dashed line is
the result of a calculation by Blankenbecler and Drell\cite{dick},
normalized to our Bethe-Heitler Monte Carlo.  Because Blankenbecler
and Drell do not include dielectric suppression or transition
radiation in their calculations, the calculations are suspect below 5
MeV (1.5~MeV) at 25 (8)~GeV.  At 25~GeV, Blankenbecler and Drell are
an excellent fit to the data, with a $\chi^2/DOF$ of 1.15 above 2 MeV.
At 8~GeV, the agreement is not as good, with $\chi^2/DOF$=2.3. Because
of the more gradual onset of suppression in the Blankenbecler and
Drell calculation, the downturn in the 8~GeV spectrum occurs above $k=500$
MeV and is not visible.  

At 25~GeV, the prediction of Shul'ga and Fomin is shown as a straight
dot-dashed line.  At 8~GeV, the target is thin enough that their
formulae do not apply. Zakharov\cite{Zakharov} has compared his
calculation with our 0.7\% \xo 25~GeV data for $k> 5 $MeV, and finds
excellent agreement.

Figure~13 shows data from the 0.1\% \xo gold target, with
Bethe-Heitler (dashed line) and dielectric suppression only (solid
line) Monte Carlos.  The target is thin enough that the total multiple
scattering is less than $1/\gamma$.  One might expect that there is
then no LPM suppression. However, Blankenbecler and Drell
found\cite{dick} a slight suppression at 25~GeV, about 8\% at $k=500$
MeV, rising to 13\% at $k=100$ MeV.  At 8~GeV, the suppression is a
few percent.  Because of the small signal and relatively large
uncertainties, we are not able to confirm or reject this slope.

Little transition radiation is visible.  Because $t<l_f$ over the
entire relevant $k$ range, transition radiation is reduced by
$\sin^2{(t/l_f)}$\cite{Artru} compared to a thick target ($t>l_f$).
Dielectric suppression is expected to be similarly reduced, because
the total phase shift in the entire target thickness is much less than
one.  However, at 8~GeV, considerable downturn is observed, with the data
between the dielectric suppression only and Bethe-Heitler predictions.

Unfortunately, there are a number of experimental uncertainties
associated with this target.  Because the target is so thin,
background contamination is relatively more significant than it is for
other targets.  The actual target thickness is not well known, and
visual inspection suggests that the target thickness is not uniform;
we have not been able to measure this.  We have observed considerable
variation in overall bremsstrahlung amplitude from run to run; this
could be caused by the beam spot hitting different locations on the
target.

\section{Target Subtraction}

The data presented above show that the suppressed curves are a much
better fit to the data than the Bethe Heitler curves.  However, in
many cases, the Monte Carlo does not fit the data well, especially
when the target thickness is a significant fraction of $l_f$, and
surface effects are large.  One way to remove the surface effects is
to compare targets of the same material, but differing thicknesses.

We do this by performing a bin by bin subtraction of the histograms of
the same material but differing thicknesses, for example 6\% $X_0$ Au
- 0.7\% $X_0$ Au, giving the `middle' 5.3\% $X_0$ of the target.
Because this subtraction increases the slope change due to
multi-photon pileup (multiple interactions in the target), it is
necessary to compare the result with Monte Carlo data which have been
subjected to the same procedure.  The subtractions are shown in
Figs.~5--11.

This subtraction suffers from a few drawbacks. It assumes that the
target is thicker than a formation length, so that there is no
interference between the transition radiation from the two edges.  The
subtraction increases the effect of multi-photon emission and photon
absorption in the targets.  Because of this, when the procedure is
applied to Monte Carlo data, the result is negative below about 1 MeV
(500 keV) at 25 (8)~GeV beam energy, depending on the target material. These
effects are included in the Monte Carlo, but the subtractions do
increase the relative systematic errors.  However, edge effects change
the multi-photon pileup slightly.  Because this is not in the Monte
Carlo, it also adds to the systematic errors. The systematic errors
due to the Monte Carlo in Table~IV should be doubled.  Nevertheless,
subtraction appears to be an effective process for separating edge
effects from bulk LPM suppression, so we present the subtracted data
here.

After subtraction, the LPM Monte Carlo is a much better match to the
data.  To quantify the agreement, we have performed a $\chi^2$ fit of
the Monte Carlo to the data; the results of the fit are given in
Table~III.  The only free parameters in the fit are the previously
mentioned normalization constants; see Section VIII for a discussion
of the normalization.  For most of the materials, the fit quality is
good, with $\chi^2$/DOF$\sim 1$.  For the targets where the
$\chi^2/DOF > 1$, indicating a poor fit, the disagreement appears to
be within the systematic errors; we have not attempted to include the
systematic errors in the fit or $\chi^2$.

Figures~5c and 5f show the carbon data, above 450 keV (200 keV) for 25
(8)~GeV.  The fit quality is reasonable, although, because of the good
statistics, the $\chi^2/DOF$s at 25~GeV of 2.74 is high.  At 8~GeV the
fit quality is much better, with $\chi^2/DOF=1.17$.  At 25 GeV, much
of the $\chi^2$ comes from the region of small $k$, where the data are
below the Monte Carlo.

The fact that the subtracted data and MC agree much better than their
unsubtracted counterparts indicates that the mismatch between the data
and LPM + dielectric suppression MC is related to the target
edges. This is a bit puzzling, since it is difficult to see how
surface terms could {\it increase} the suppression; an unreasonably
large contamination by a higher $Z$ material would be required to
explain the spectrum.

Figures~6d, 7d and 9d show the 25~GeV subtracted aluminum, iron and
tungsten data, above 500 keV.  The aluminum and tungsten simulations
are an excellent fit to the data, with $\chi^2/DOF$=0.84 and 0.99
respectively.  The iron fit is rather poor with $\chi^2/DOF= 2.32$,
although it agrees a lot better than the unsubtracted data.  

Figures~10c and 10f show the uranium data, above 1000 keV (300 keV) for
25 (8)~GeV.  The fit quality is quite good, with $\chi^2/DOF$s of 0.89
and 1.56.

Figures~11c and 11f show the gold data, above 5 MeV (350 keV) for 25
(8)~GeV.  The fit quality is excellent at 25~GeV, with
$\chi^2/DOF$=0.85.  The 8~GeV data have a $\chi^2/DOF$ of 2.68,
because the data are below the MC prediction below 1 MeV.  This may be
partly because the 0.7\% \xo target is so thin that coherent
interactions between the two edges are significant.  However, in that
case we would expect better agreement at 8~GeV, where $l_f$ is much
smaller.

One side benefit of the subtraction procedure is that the break in the
spectrum between LPM suppression and dielectric suppression becomes
much clearer.

>From these results, it is clear that the Migdal formula does an
excellent job of describing suppression in bulk media.  The
suppression scales as expected with beam energy, photon energy, target
Z and $X_0$.

It would be possible to modify the subtraction procedure to isolate
the emission due to a single edge.  However, because of the large
errors and uncertainties inherent in the process, the results would
have limited significance. For the carbon and iron targets, the `edge'
term would be negative over a fair fraction of the spectrum.

\section{Systematic Errors}

Our systematic errors are divided into two classes, those that affect
the absolute normalization only (discussed in the next section) and those
that can affect the shape of the spectrum.   The major systematic errors
are due to energy calibration, photon (cluster) finding, calorimeter 
nonlinearity, uncertainty in the target density, and multiphoton pileup,
as summarized in Table~IV.  

The systematic errors that can affect the spectral shape are quite
different for the high and low calorimeter gain data, because several
things change.  As was previously discussed, in one case energy loss
is primarily by showering, and in the other by Compton scattering, so
the clustering works differently.  Also, for the high gain data,
backgrounds are much larger.  For these reasons, the systematic errors
are much larger for $k<5$~MeV than for $k>5$~MeV.  Surprisingly, except
for the synchrotron radiation removal cut, the systematic errors are
independent of electron beam energy.

For $k>5$ MeV, the major errors are calorimeter energy calibration
(1.5\%), photon cluster finding (2\%), calorimeter nonlinearity (3\%),
backgrounds (1\%), target density(2\%), electron flux (0.5\%), and
Monte Carlo uncertainties (1\%), for a total systematic uncertainty
of 4.6\%.

The 5\% uncertainty in the calorimeter energy calibration is
equivalent to shifting the histogrammed data by just over half a bin.
The magnitude of the consequent error in cross section depends on the
slope of the curve, and consequently on the target thickness.  In the
worst case, the 6\% $X_0$ gold target, a 5\% energy scale shift
produces a 1.5\% change in the measured cross section.

The photon cluster finding introduces a 2\% uncertainty in the cross
section.  Likewise, leakage out the back and sides of the calorimeter,
and PMT saturation effects introduces a 3\% uncertainty,

Most of the targets materials had a well defined density.  However,
the carbon targets were graphite, which has a density that can vary,
only partly because it can absorb water.  During data taking, they were in
vacuum, so that wasn't a problem.  Their density was determined by
measuring and weighing them, the latter after they were dried in an
oven.  We measured a density 4$\pm$2\% below the standard
value\cite{PDG}, and used this density in calculations of the
radiation length and $E_{LPM}$.

For the $k<5$ MeV data, many systematic errors are larger.  The photon
cluster finder is less effective because of the possibility of
non-contiguous energy deposition (7\%), and the calorimeter energy
calibration is worse due to the need to use the higher energy data as
an intermediate calibration (3\%).  Also, at these energies,
backgrounds are larger, a 4\% uncertainty, and the Monte Carlo is
probably less accurate for low energy photons (1.5\%) This gives an
overall 9\% systematic error.

For the data where the synchrotron radiation rejection cut was
used, `high gain' 25~GeV running, there is an additional systematic
error.  This is because the cut efficiency is sensitive to how well
the electron beam is centered on the calorimeter.  During our running,
the average deviation from the calorimeter center was less than 5 mm.
This introduces an additional 15\% systematic error.  

\section{Normalization} 

We have compared our measured absolute cross sections with the Migdal
predictions by calculating the adjustment required to normalize the
data to the Migdal plus dielectric suppression Monte Carlo.  To avoid
regions where edge effects and backgrounds are important, the 25~GeV
data are normalized over the range 20 MeV to 500 MeV, and the 8~GeV
data are normalized from 2 MeV to 500 MeV.  For the 0.7\% $X_0$ data, a
narrower range, 30~MeV (10 MeV) to 500 MeV was used at 25 (8)~GeV, to
avoid surface effects.  This is a much wider fitting range than was
used previously\cite{prl}. For each data set, Table~II gives the
normalization corrections, the percentage by which it is necessary to
adjust the Monte Carlo prediction to best match the data.  The errors
given are statistical only; the systematic errors are summarized in
Table~IV.

The electron flux was measured using the lead glass blocks.  The
blocks are large enough so that there was almost no leakage out the
side or top of the block stack.  The major source of missed electrons
was high energy bremsstrahlung where the electron lost enough energy
to be bent below the lead glass blocks.  Electrons with energies below
17.4 (5.8)~GeV for 25~(8)~GeV beams missed the blocks.

The fraction of electrons missing the blocks depended on the target
thickness, and was determined by the Monte Carlo; the miss probability
ranged from 2\% to 7\%. This miss probability was folded into a matrix
to estimate the number of single electron events.  Because missed
electrons events produce high energy photons, the events will also
cause overflows in the calorimeter, thus they do not affect the
histograms.

In this unfolding, a fortuitous cancellation limits the systematic
errors to 0.5\%.  Most of our running was at an average of one
electron per pulse.  At this level, the probability of a single
electron being missed was very close to the probability of a two
electron event appearing as a single electron in the lead glass
blocks.  So, the probability of losing an electron almost completely
cancels out of the luminosity, so it is not necessary to know this
number well.

Many uncertainties that affect the relative measurement are reduced
for the normalization, because of the more limited photon energy
range.  Above 20 MeV, photon-finding is much more robust, and the
calorimeter nonlinearities are less significant.

The target thickness measurement was more complicated than originally
expected.  The targets thicknesses were measured with calipers.  The
thinner targets were weighed, and their sizes measured, to find the
thickness in gm/cm$^2$.  The uncertainty in thickness contributed a
2\% systematic error. Because of the previously mentioned
uncertainties about the 0.1\% $X_0$ gold target, it is not considered
here.

The normalization constant depends only slightly on the
normalization procedure.  Changing the lower energy limit only
produces small changes, of order 0.5\%. To account for these fitting
uncertainties, we include a 1\% systematic error.

On the average, the normalizations show that the data are slightly
below the Migdal prediction.  The weighted averages are $-4.7\pm2.0$\%
($-3.1\pm 5.6$\%) at 25 (8)~GeV, with a 3.5\% systematic error.  If
the outlying 6\% $X_0$ gold target is excluded from the 8~GeV data,
the average becomes $-4.8\pm2.5$\%.  However, the 2.5\% contribution
to the cross section from the $(1-y)(Z^2+Z)/3$ term discussed in
section II.A. increases the disagreement.  Including systematic errors,
we find roughly a $2\sigma$ discrepancy.  This is difficult to explain
by experimental effects alone.

There are some attractive theoretical explanations, stemming from
limitations in Migdal's calculations.  Migdal used a Gaussian
approximation for multiple scattering.  This underestimates the
probability of large angle scatters.  These occasional large angle
scatters would produce some suppression for $k>k_{LPM}$, where Migdal
predicts no suppression and where we determine the normalization.
Fig.~12 shows that, compared to Migdal, the suppression predicted by
Blankenbecler and Drell turns on much more slowly, and, hence if
Blankenbecler and Drell were used in the normalization, the
discrepancy would be lessened or eliminated.  Zakharov's\cite{Zakharov}
calculation would also appear to lessen or eliminate this discrepancy.

\section{Discussion}

As the data presented above shows, the LPM and dielectric
effects suppress bremsstrahlung as expected for most of our target
materials and thicknesses.  The suppression scales as expected with
electron energy, photon energy, and radiation length. Materials with
similar radiation lengths, but different densities and atomic numbers
(tungsten and uranium) display similar LPM suppression. For low photon
energies, the formation length can become longer than the target
thickness.  When that happens, we observe that the target behaves as a
single scatterer, and the spectrum again becomes flat, like the
Bethe-Heitler result, but at a lower intensity.  For thicker targets,
there is an edge effect radiation which can be removed by subtraction.

Unfortunately, we have not found a single calculation that matches the data
and includes both LPM and dielectric suppression for finite target
thicknesses. However, we have removed the finite target thickness
effects by subtraction.

Although the data clearly demonstrate LPM suppression to good
accuracy, for low $Z$ targets, the simulations do not match the data
as well as expected.  The fact that the discrepancy is greatly reduced
by the subtraction procedure indicates that some sort of a surface
effect is involved.  However, it is difficult to imagine how a surface
effect can reduce the emission.  It is also difficult to imagine
instrumental effects that would affect only carbon and iron; a 20\%
adjustment to the the energy scale would improve the agreement for
these materials, but it would produce a large disagreement for the
other materials.

A discrepancy in the bulk material (subtracted plots) might be
explainable by material effects.  The carbon targets were made of
pyrolitic graphite, which has internal structure on a scale much
larger than crystalline structure.  If the target varied in density on
a scale large with respect to the formation zone length, then the
average suppression and edge effects will increase and additional
transition radiation will be generated, consistent with the data at
25~GeV beam energy.

The iron targets should be mechanically homogeneous, but magnetically
inhomogeneous. Individual magnetic domains are magnetized to
saturation ($B\sim 2T$), but in different directions.  The typical
domain size is of order 1$\mu$m.  Magnetic bending of the elctrons can
also suppress bremsstrahlung; a detailed model of the phenomenon is
lacking\cite{klein}.  Two Tesla is enough to bend the electrons by
$1/\gamma$ in a distance $l_f$; in combination with multiple
scattering, this could alter the spectrum. 

It is perhaps significant that at 8~GeV beam energy, where the formation
zone is a factor of 10 shorter and edge effects consequently are greatly
reduced, the data shows  much better agreement than at 25~GeV beam energy.

\section{Conclusions}

The LPM and dielectric effects suppress bremsstrahlung as expected for
a variety of target materials and thicknesses and two beam energies.
For carbon and iron, somewhat more suppression than expected is
observed. However, the excess suppression appears to be a surface or
magnetic effect, and perhaps can be explained by the 
properties of these targets.  For most of our targets, the agreement
is within 5\% of the theory.

Thin targets, where the formation length is longer than the target
thickness, behave as single radiators.  Calculations by Blankenbecler
and Drell reproduce the shape of the photon spectra where dielectric
suppression is unimportant.  

The overall bremsstrahlung cross section for low energy photons is
measured to be about 5\% (2$\sigma$) lower than expected due to
Migdal's work.  Alternate calculations, by Blankenbecler and Drell, or
by Zakharov, might agree better with the data.

\section{Acknowledgements}
We would like to thank the SLAC Experimental Facilities group for
their assistance in setting up the experiment, the SLAC Accelerator
Operations group for their efficient beam delivery, and the SLAC
Computing Services group for providing the data analysis facilities.
We also acknowledge useful conversations and cross section
calculations from Sid Drell and Richard Blankenbecler.  N. Shul'ga and
S. Fomin explained the details of their calculatons to us. Don Coyne
provided much direct and indirect support.  This work was supported by
Department of Energy contracts DE-AC03-76SF00515 (SLAC),
DE-AC03-76SF00098 (LBNL), and National Science Foundation grants
NSF-PHY-9113428 (UCSC) and NSF-PHY-9114958 (American U.).

\vfill\newpage
\begin{figure}
\vspace*{6.5in} \hspace*{.45in}
\includegraphics{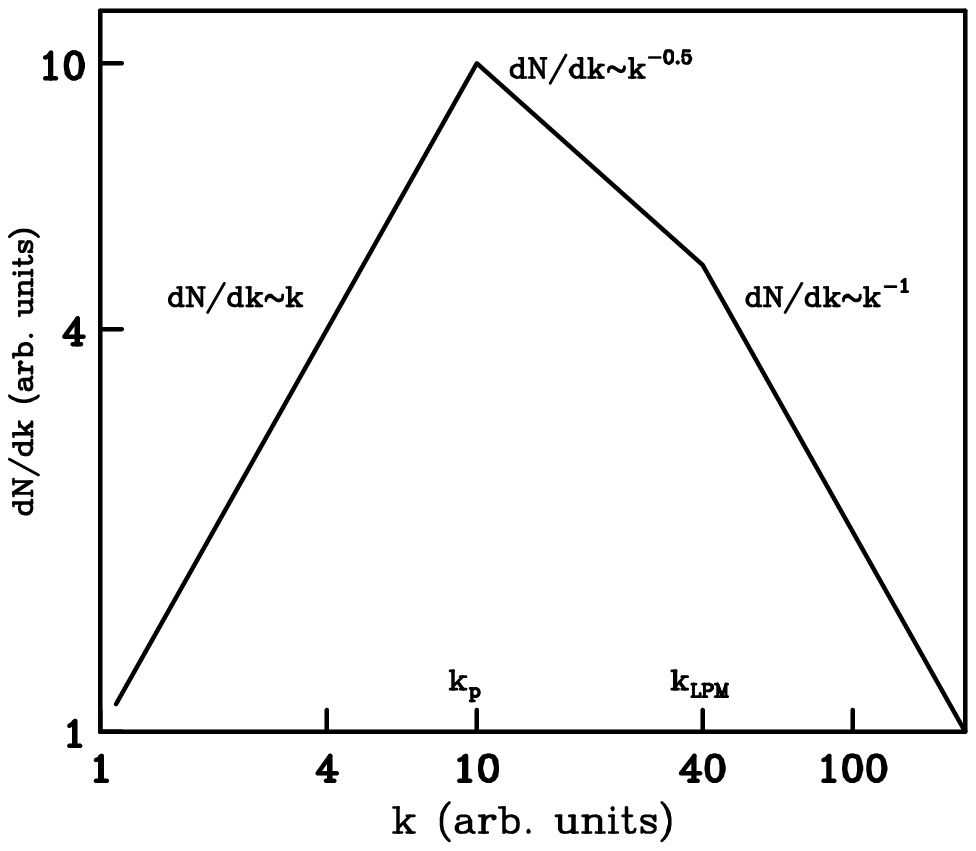}
\caption{Schematic plot of cross sections, showing Bethe-Heitler, LPM
and dielectric suppression regions.}
\end{figure}

\begin{figure}
\vspace*{6.5in} \hspace*{.45in}
\includegraphics{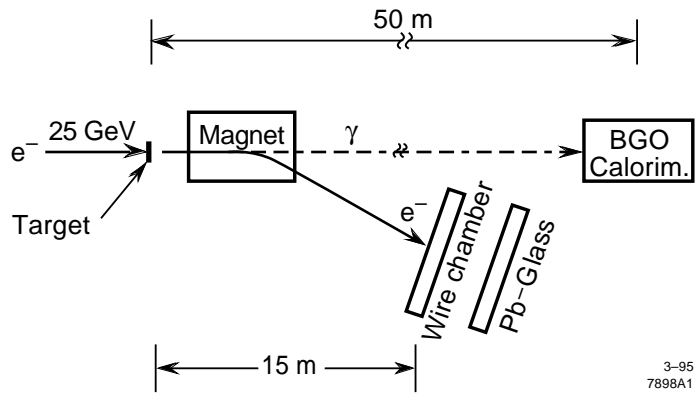}
\caption{A diagram of the experiment.  The
apparatus is described in detail in the text.}
\end{figure}

\begin{figure}
\vspace*{6.5in} \hspace*{.45in}
\includegraphics{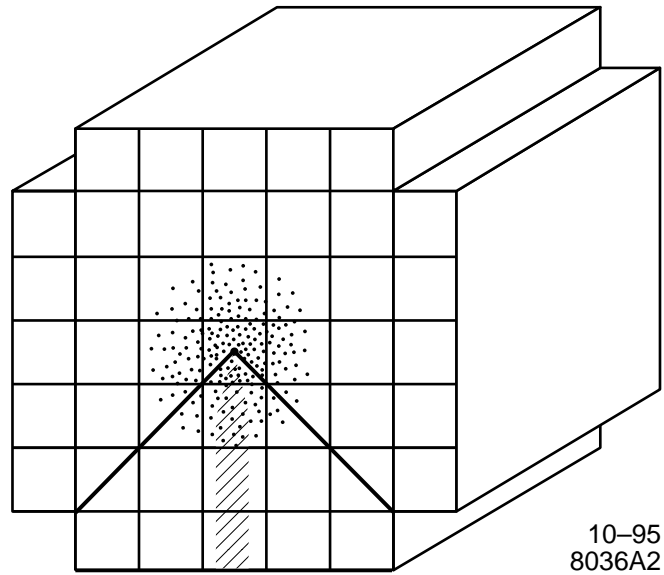}
\caption{Drawing of the front of the calorimeter, showing
bremsstrahlung plus transition radiation signal (dots) and synchrotron
radiation background (hashes).  The region below the diagonal solid
lines is where the background rejection cut removes photon clusters
from the data.}
\end{figure}

\begin{figure}
\vspace*{6.5in} \hspace*{.45in}
\includegraphics{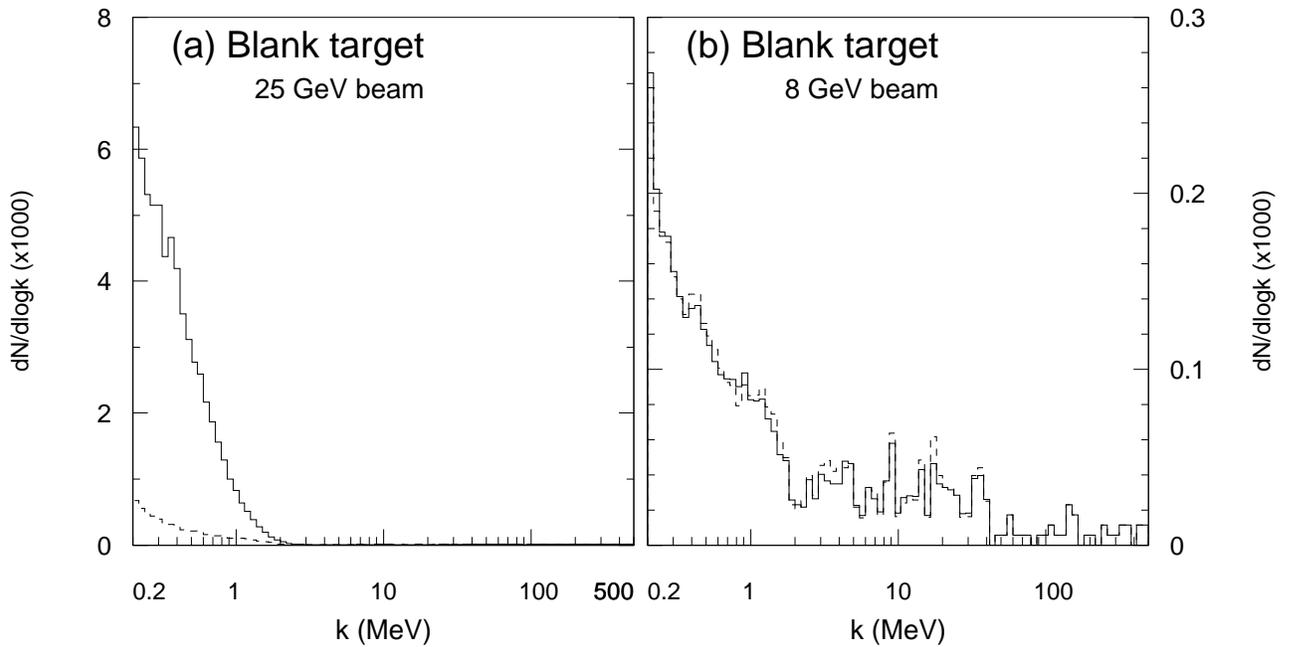}
\caption{Data from the blank target runs at (a) 25~GeV and (b) 8~GeV.
The units are photons per log($k$) per 1000 electrons.  There are 25
bins per decade of photon energy, so each bin has width $\Delta
k/k\sim 0.09$.  The raw data are shown in the solid histogram, while
the dashed lines show the data after the synchrotron radiation removal
cut and efficiency correction.  The cut removes about 90\% of the data
at 25~GeV, while at 8~GeV the efficiency only matches the geometrical
expectation.}
\end{figure}

\begin{figure}
\vspace*{5.5in} \hspace*{.45in}
\includegraphics{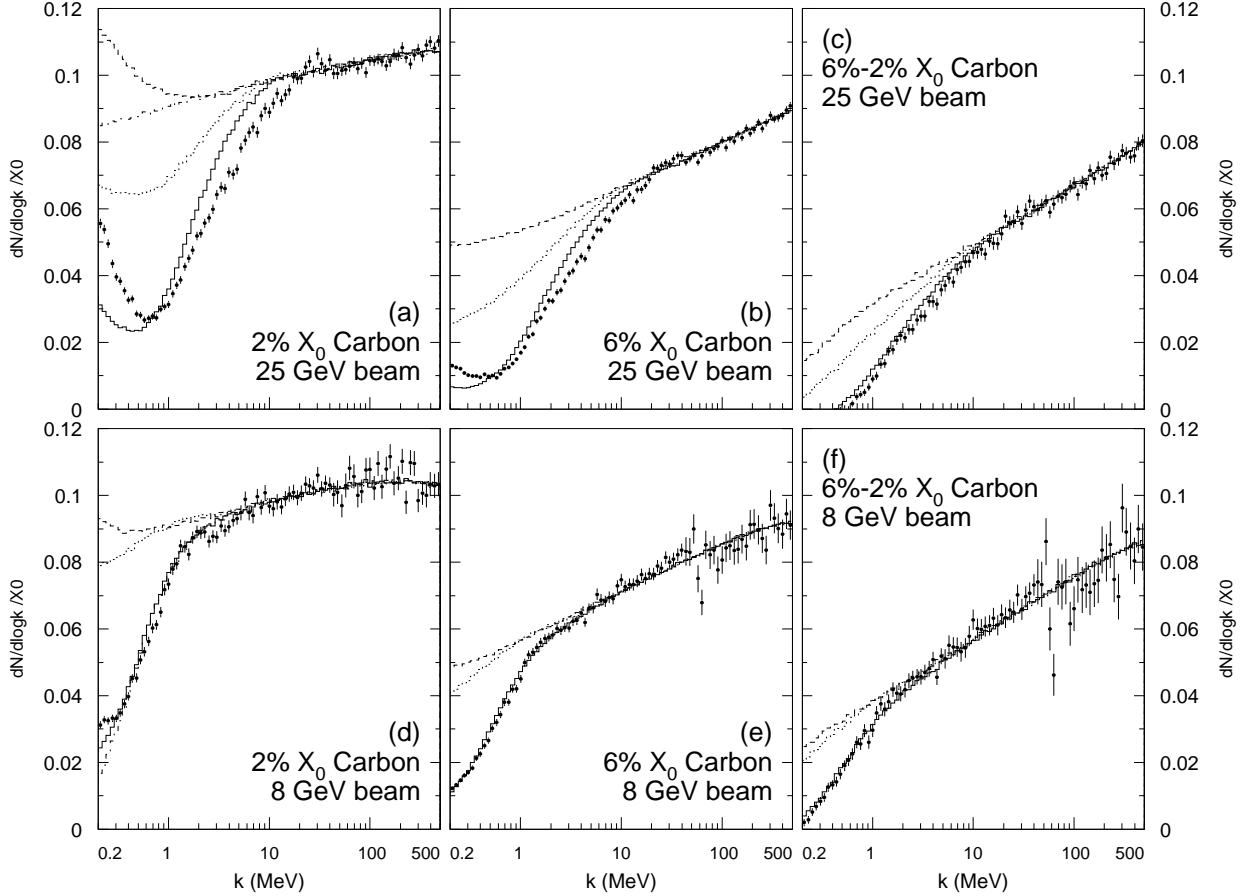}
\caption{Measurements with statistical errors only of $dN/d(\log k)$
compared with the LPM plus dielectric effect plus transition radiation
Monte Carlo curves (solid line), for our (a) 2\% \xo carbon and (b)
6\% \xo carbon targets in 25~GeV electron beams, while (d) shows the
2\% \xo carbon and (e) the 6\% \xo carbon target in the 8~GeV beam.
The cross sections are given as $dN/d(\log k)/X_0$ where $N$ is the
number of events per photon energy bin per incident electron.  (c)
shows the result of subtracting the data in (b) from that in (a),
while (f) is the result of subtracting (e) from (d), as discussed in
Section VI of the text.  The curves are cut off where they go negative
as a result of the procedure.  Also shown are the Bethe-Heitler plus
transition radiation MC (dashed line), LPM suppression only plus
transition radiation (dotted line) and, for comparison, Bethe-Heitler
without transition radiation (dot-dashed line).  Figures 6 through 12
follow a similar format.}
\end{figure}

\begin{figure}
\vspace*{6.5in} \hspace*{.45in}
\includegraphics{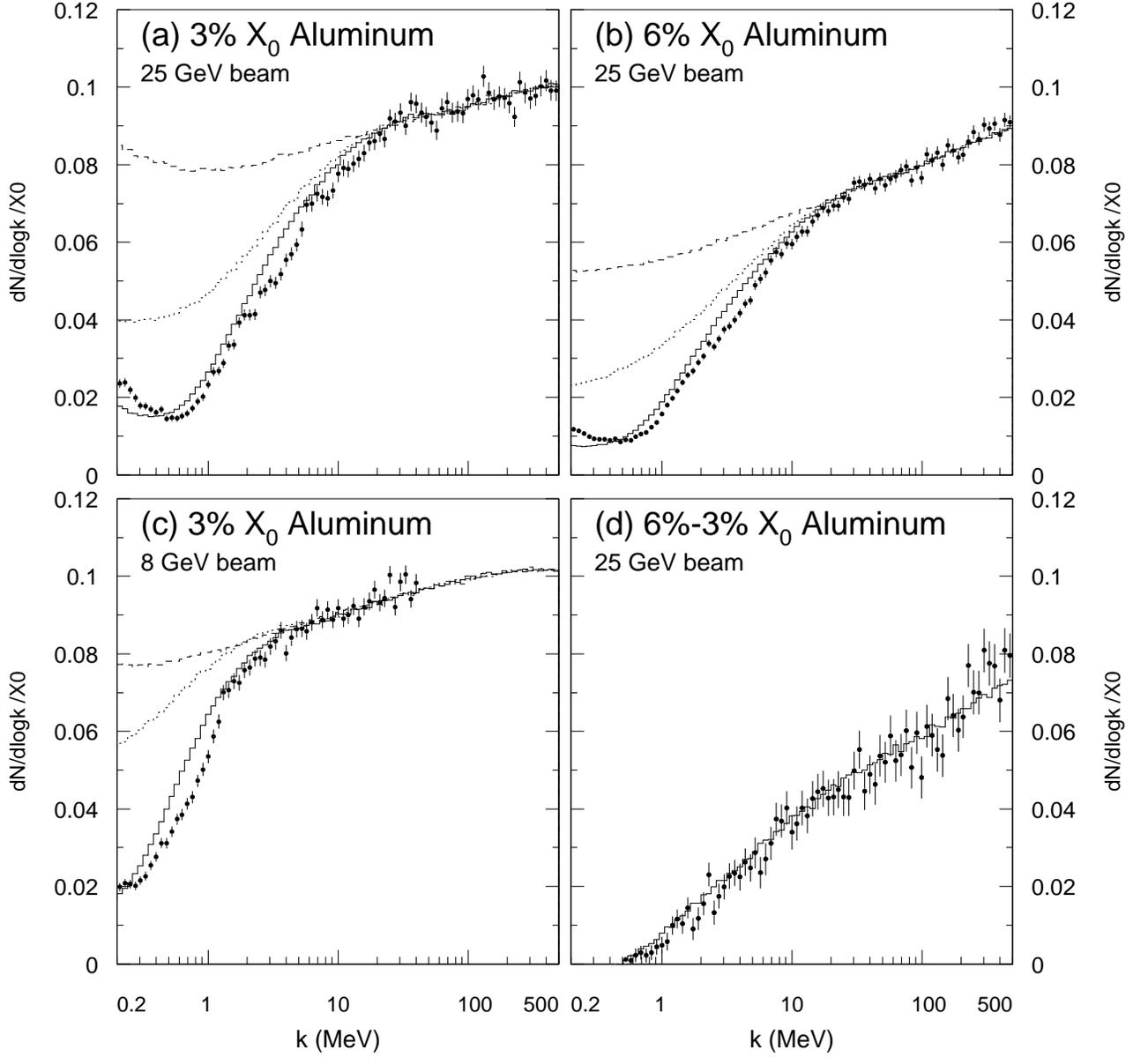}
\caption{Measurements and Monte Carlo for our (a) 3\% \xo and (b) 6\%
\xo aluminum targets at 25~GeV and (c) 3\% \xo at 8~GeV.  (d) is the
result of subtracting (b) from (a). The data and Monte Carlo formats
and labels match Fig. 5, except that, for variety, the LPM suppression only
simulation does not have transition radiation added.}
\end{figure}

\begin{figure}
\vspace*{6.5in} \hspace*{.45in}
\includegraphics{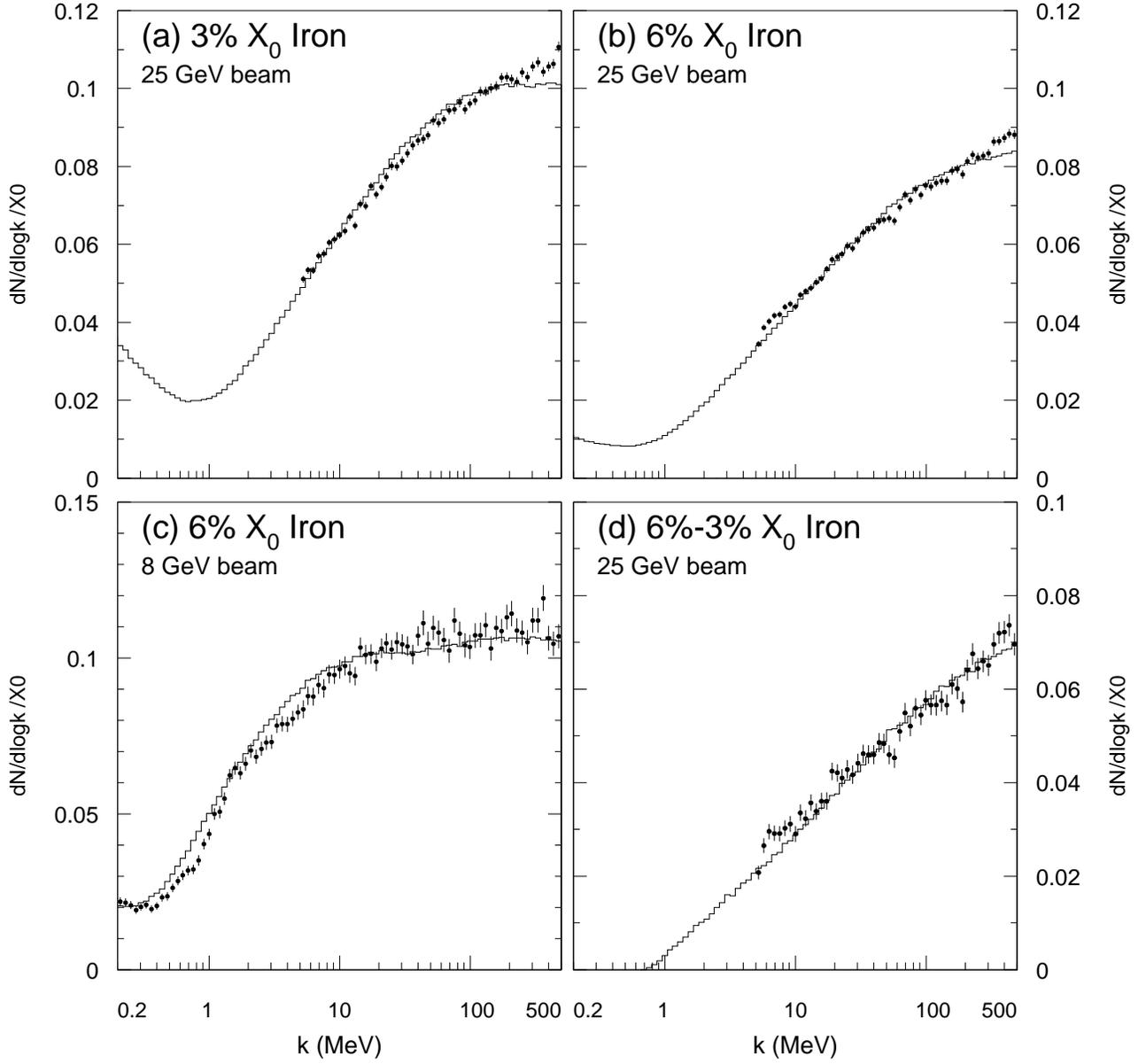}
\caption{Measurements and Monte Carlo for our (a) 3\% \xo and (b) 6\%
\xo iron targets at 25~GeV and (c) 6\% \xo at 8~GeV, while (d) is
the result of subtracting (b) from (a).  Here, only an LPM
plus dielectric suppression curve is shown.}
\end{figure}

\begin{figure}
\vspace*{6.5in} \hspace*{.45in}
\includegraphics{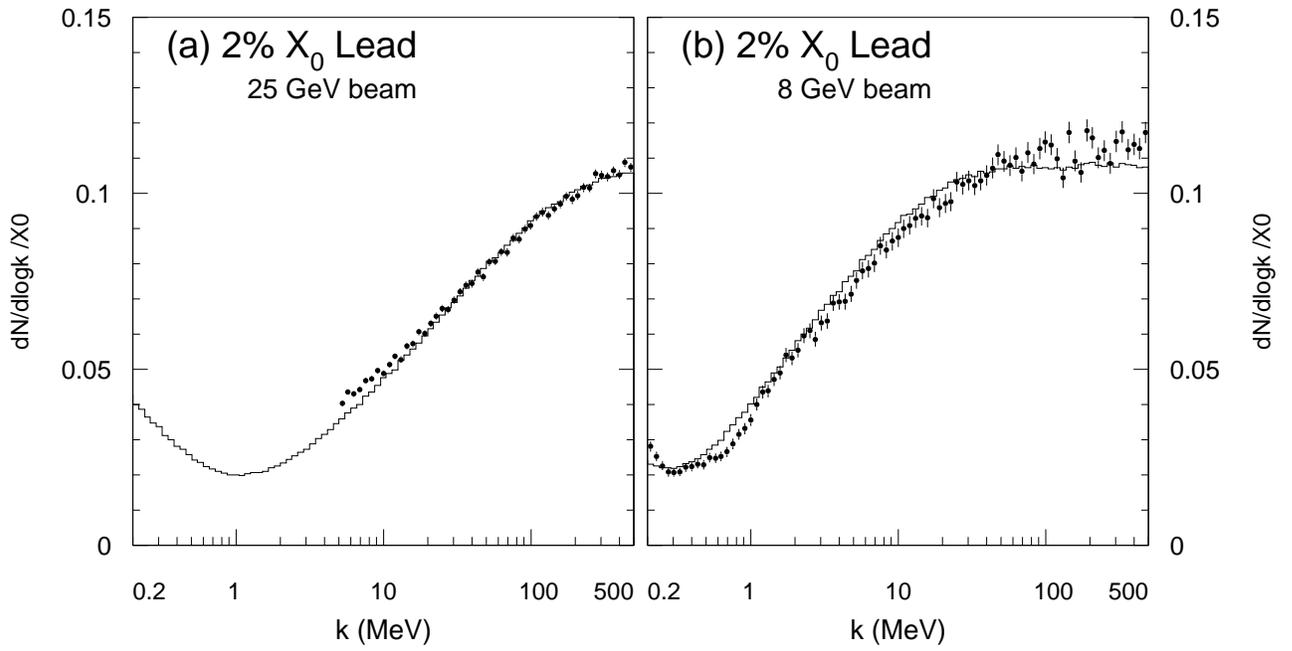}
\caption{Measurements and Monte Carlo for our 2\% \xo lead target at
(a) 25~GeV and (b) 8~GeV.}
\end{figure}

\begin{figure}
\vspace*{6.5in} \hspace*{.45in}
\includegraphics{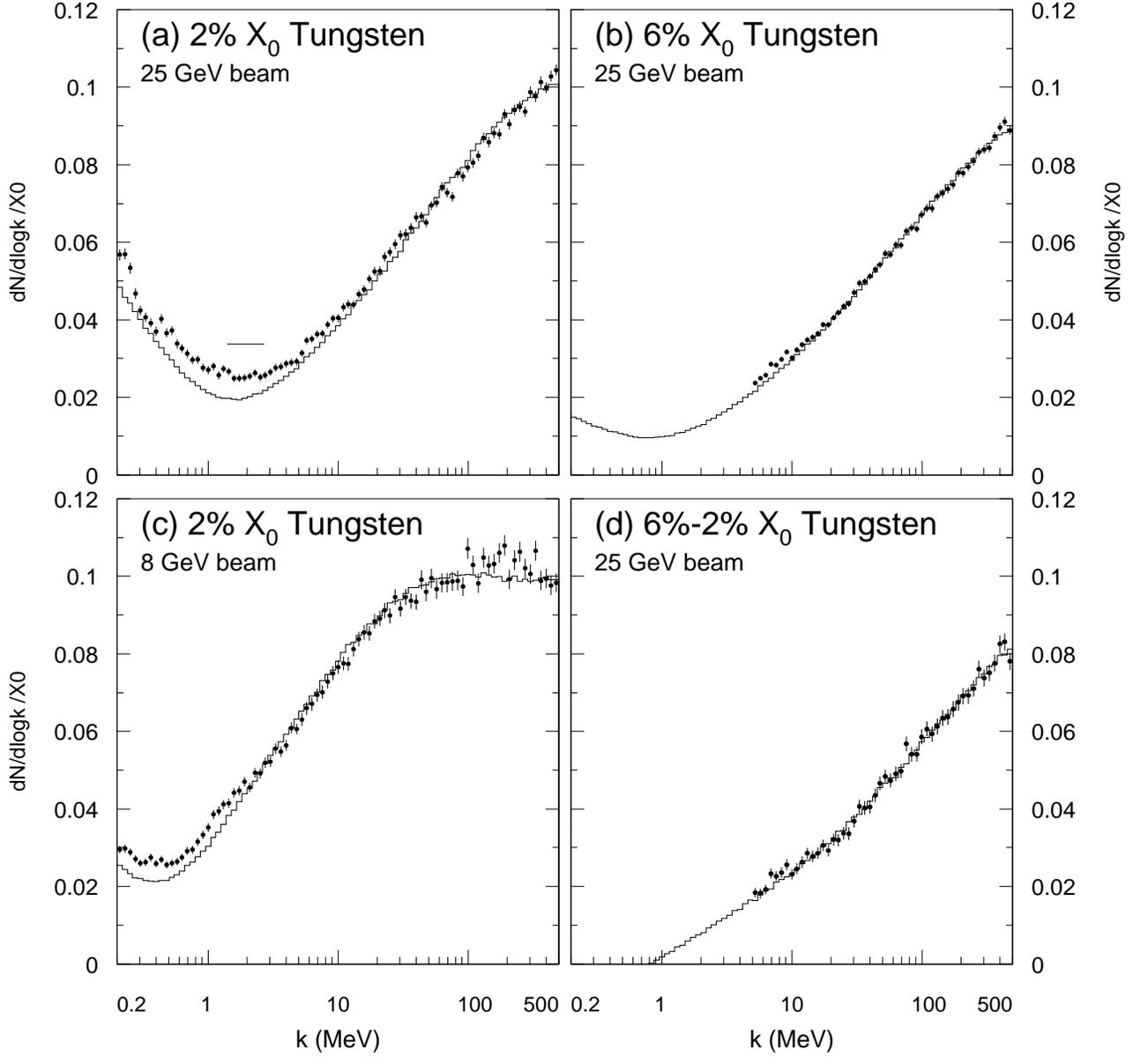}
\caption{Measurements and Monte Carlo for our (a) 2\% \xo and (b) 6\%
\xo tungsten targets at 25~GeV and (c) 2\% \xo at 8~GeV, while (d) is
the result of subtracting (b) from (a).  The flattening below 10 MeV
is discussed in the text. The straight solid line in (a) between 1.5
and 2.7 MeV is the `single radiator' calculation.}
\end{figure}

\begin{figure}
\vspace*{6.5in} \hspace*{.45in}
\includegraphics{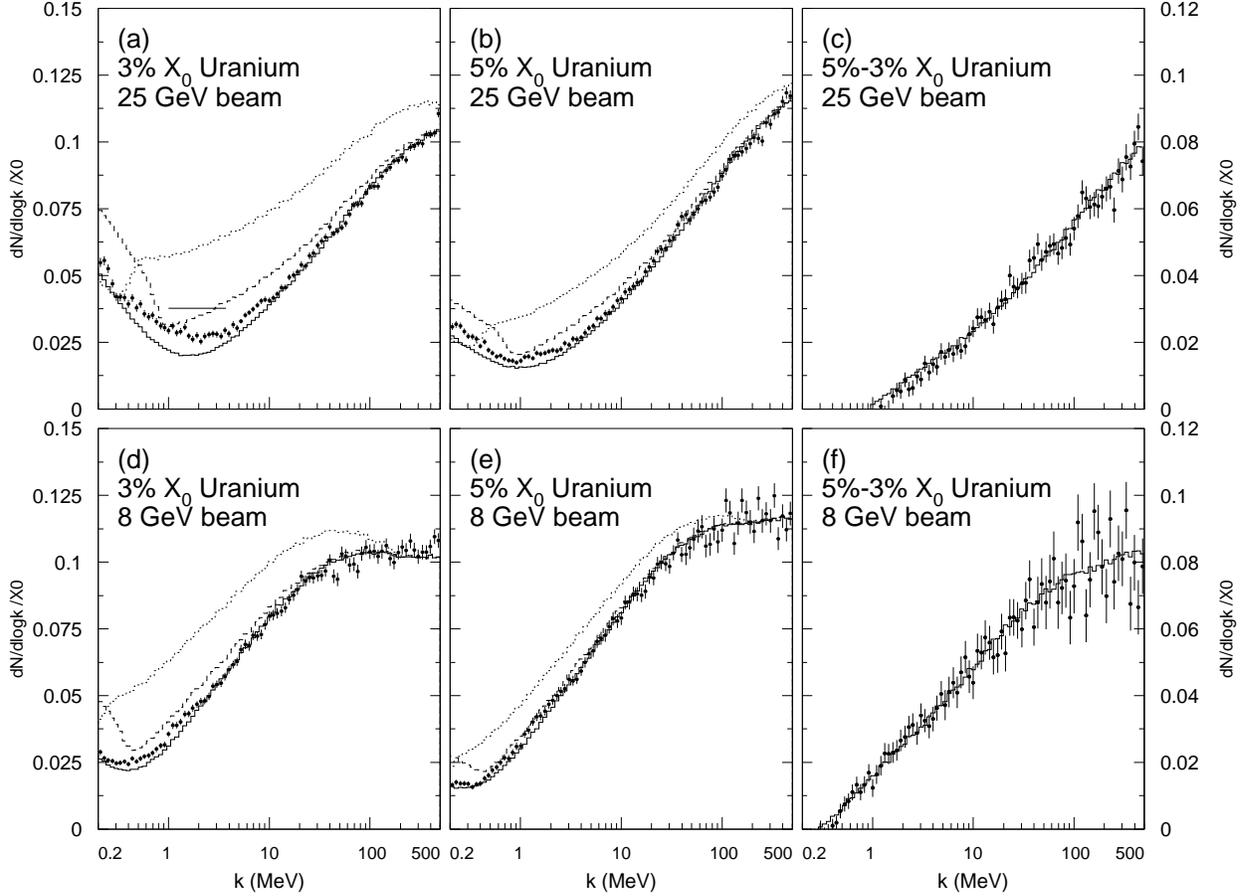}
\caption{Measurements and Monte Carlo for our (a) 3\% \xo and (b) 5\%
\xo uranium targets at 25~GeV and (d) 3\% \xo and (e) 5\% \xo uranium
at 8~GeV, while (c) and (f) are the subtracted data.  The solid line
shows the standard Monte Carlo prediction. The horizontal solid line 
in (a) is from Eqn.~14. The dashed line includes
the Pafomov transition radiation in the Monte Carlo, along with LPM
and dielectric suppression.  The dotted line is for the LPM and
dielectric suppression plus the Ternovskii transition radiation with
$\chi=1$.}
\end{figure}

\begin{figure}
\vspace*{6.5in} \hspace*{.45in}
\includegraphics{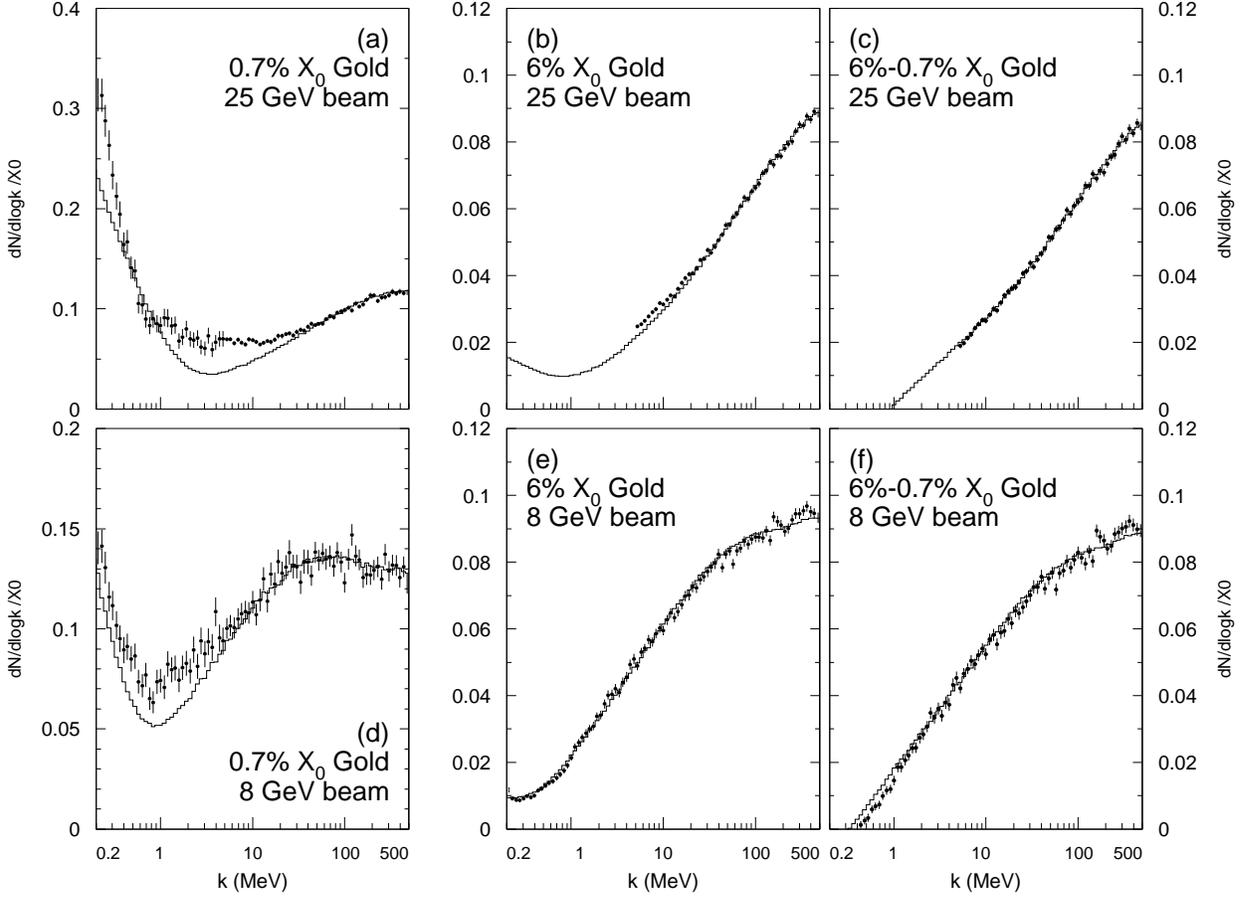}
\caption{Measurements and Monte Carlo for our (a) 0.7\% \xo and (b)
6\% \xo gold targets at 25~GeV and (d) 0.7\% \xo and (e) 6\% \xo gold
at 8~GeV, with the subtracted data shown in (c) and (f).}
\end{figure}

\begin{figure}
\vspace*{6.5in} \hspace*{.45in}
\includegraphics{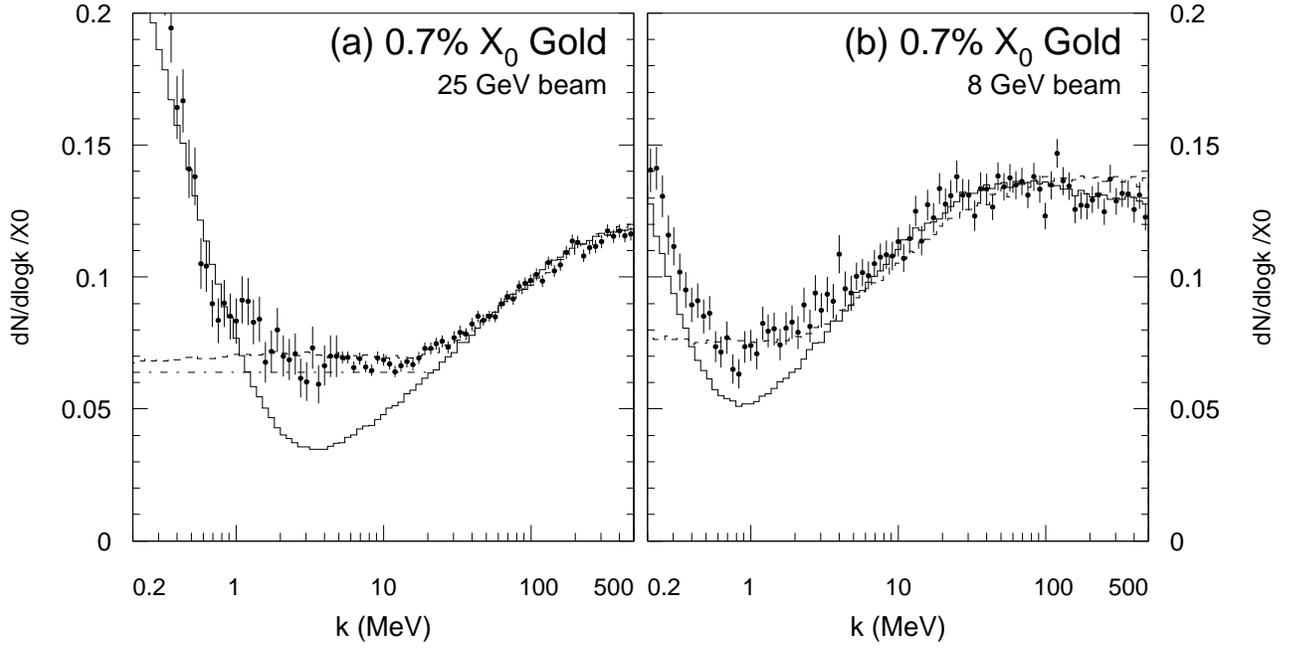}
\caption{Expanded view of the 0.7\% $X_0$ data at (a) 25~GeV and (b)
8~GeV, compared with the Blankenbecler and Drell prediction (dashed
line), Shulga and Fomin prediction (dot-dashed line). For comparison,
the standard MC is shown as the usual solid line.}
\end{figure}

\begin{figure}
\vspace*{6.5in} \hspace*{.45in}
\includegraphics{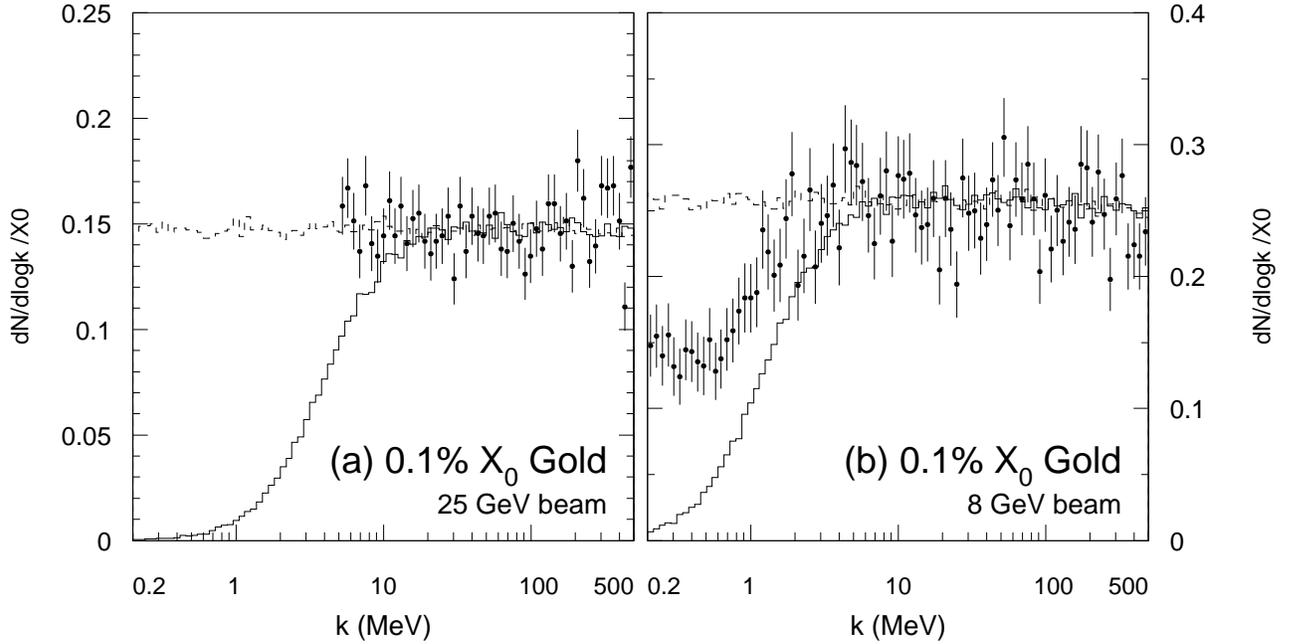}
\caption{Measurements and Monte Carlo for our 0.1\% \xo gold target at
(a) 25~GeV and (b) 8~GeV.  The dashed line is the Bethe-Heitler prediction (no
suppression), while the solid line is a Monte Carlo which includes
dielectric suppression, but not LPM suppression. Because the very thin
target should exhibit little transition radiation, no transition
radiation is included in the Monte Carlo.}
\end{figure}

\begin{table}[t]
\caption{$E_{LPM}$, $k_{LPM25}$, $k_{LPM8}$ and $r$ for
the target materials used here.}
\begin{tabular}{lrrrrrr}
Target & Z & $X_0$ (cm) &
$E_{LPM}$ (TeV) &  $k_{LPM25}$ (MeV) & $k_{LPM8} $(MeV)  
& $r$\ \ \  \\
\hline
Carbon & 6& 19.6 & 74  &  8.5 & 0.87 & $5.5\times10^{-5}$ \\
Aluminum &13 & 8.9 & 36  &  15.7 & 1.6 & $6.0\times10^{-5}$ \\
Iron & 26 & 1.76 & 6.6  &  95 & 9.7 & $1.0\times10^{-4}$ \\
Lead & 82 & 0.56 & 2.1  &  295 & 30.1 & $1.1\times10^{-4}$ \\
Tungsten & 74& 0.35 & 1.32  &  472 & 48.3 & $1.5\times10^{-4}$ \\
Uranium  &92 & 0.35 & 1.32  &  472 & 48.3 & $1.4\times10^{-4}$ \\
Gold     &79 & 0.33 & 1.25  &  500 & 51.2 & $1.5\times10^{-4}$ \\
\end{tabular}
\end{table}

\begin{table}[t]
\caption{List of target thicknesses and overall normalization
constants.  The target thicknesses $t$ are given in mm, gm/cm$^2$, and
$X_0$.  The last two columns give the normalization adjustments used
to match the simulations with the data (statistical errors only).}

\begin{tabular}{crrcll} Target & $t$ & & X$_0$ & Normalization & 
Normalization \\    
-      &(mm) & (g/cm$^2$) & (\%) & (\% at 25 GeV) & (\% at 8 GeV)  \\
\hline
2\% C    & 4.10  & 0.894  & 2.1  & -3.0$\pm$0.3  & -6.0$\pm$0.4 \\
6\% C    & 11.7  & 2.55   & 6.0  & -2.9$\pm$0.2  & -4.6$\pm$0.5 \\
3\% Al   & 3.12  & 0.842  & 3.5  & -2.7$\pm$0.4  & -3.0$\pm$0.4 \\
6\% Al   & 5.3   & 1.4    & 6.0  & -2.8$\pm$0.3  &  \\
3\% Fe   & 0.49  & 0.39   & 2.8  & -5.4$\pm$0.2  & -1.4$\pm$0.4 \\
6\% Fe   & 1.08  & 0.85   & 6.1  & -7.5$\pm$0.2  &   \\
2\% Pb   & 0.15  & 0.17   & 2.7  & -4.5$\pm$0.2  &  -0.7$\pm$0.4 \\
2\% W    & 0.088 & 0.17   & 2.7  & -8.3$\pm$0.3  & -8.6$\pm$0.3 \\
6\% W    & 0.21  & 0.41   & 6.4  & -4.7$\pm$0.3  &     \\
3\% U    & 0.079 & 0.15   & 2.2  & -5.6$\pm$0.3  & -6.3$\pm$0.3 \\
5\% U    & 0.147 & 0.279  & 4.2  & -7.0$\pm$0.3  & -7.5$\pm$0.4  \\
0.1\% Au &0.0038 & 0.0073 & 0.11 &     &  \\
0.7\% Au &0.023  & 0.044  & 0.70 & -1.3$\pm$0.4  & 12.2$\pm$0.7 \\
6\% Au   &0.20   & 0.39   & 6.0  & -5.5$\pm$0.2  & -5.0$\pm$0.3  \\
\end{tabular}
\end{table}

\begin{table} [t]
\caption{$\chi^2$ per degree of freedom of the fits to the subtracted
data.  The only free parameters were the absolute normalizations of
the two individual targets.  Typically, there were about 60 degrees of
freedom. Statistical errors only were included in the fit.}
\begin{tabular}{lcc}
Material&25 GeV	& 8 GeV \\ 
\hline
Carbon	&2.74	&	1.17 \\
Aluminum&0.84	&	\\
Iron	&2.32	&	1.41 \\
Tungsten&0.99	&	\\
Uranium	&1.56	&	0.79 \\
Gold	&0.85	&	2.68 \\
\end{tabular}
\end{table}

\begin{table} [t]
\caption{Table of Systematic Errors. The absolute column refers to the
cross section for $k=500 $MeV for both 8 and 25~GeV beams.  The
relative errors for $k<5$ MeV and $k>5$ MeV also apply to both 8 and
25~GeV beams, except for the synchrotron radiation removal cut,
which is added in separately.  Uncertainties in the theoretical
calculation are not included.}
\begin{tabular}{lccc} Source 	&   & $k>5 $MeV & $k<5 $MeV \\
			& Absolute & Relative & Relative \\
\hline
Energy Calibration 	& 1\%	& 1.5\%	& 3\% \\
Photon Cluster Finding 	& 	& 2\%	& 7\% \\
Calorimeter Nonlinearity& 2\%  	& 3\% 	& 3\% \\
Backgrounds		& 	& 1\%	& 4\% \\
Target thickness	& 2\%	& 	&     \\
Target Density		& 	& 2\%	& 2\% \\
Electron Flux		& 0.5\%	& 0.5\% & 0.5\% \\
Monte Carlo		& 1.5\%	& 1\%	& 1.5\% \\
Normalization Technique & 1\%	&	&     \\
\hline
8 GeV Beam Total	& 3.5\%	& 4.6\%	& 9\% \\
Synchrotron Radiation Removal& 	& 	&15\% 	   \\
\hline
25 GeV Beam Total	& 3.5\%	& 4.6\%	& 17\%     \\
\end{tabular}
\end{table}

\end{document}